\documentclass[aps,prb,preprint,groupedaddress,floatfix]{revtex4}
\usepackage{graphicx,amssymb}
\begin{document}

\title{Free energy calculations along entropic pathways: II. Droplet nucleation in binary mixtures.} 
\author{Caroline Desgranges and Jerome Delhommelle}
\affiliation{Department of Chemistry, University of North Dakota, Grand Forks ND 58202}
\date{\today}

\begin{abstract}
Using molecular simulation, we study the nucleation of liquid droplets from binary mixtures and determine the free energy of nucleation along entropic pathways. To this aim, we develop the $\mu_1 \mu_2 VT-S$ method, based on the grand-canonical ensemble modeling the binary mixture, and use the entropy of the system $S$ as the reaction coordinate to drive the formation of the liquid droplet. This approach builds on the advantages of the grand-canonical ensemble, which allows for the direct calculation of the entropy of the system and lets the composition of the system free to vary throughout the nucleation process. Starting from a metastable supersaturated vapor, we are able to form a liquid droplet by gradually decreasing the value of $S$, through a series of umbrella sampling simulations, until a liquid droplet of a critical size has formed. The $\mu_1 \mu_2 VT-S$ method also allows us to calculate the free energy barrier associated with the nucleation process, to shed light on the relation between supersaturation and free energy of nucleation, and to analyze the interplay between the size of the droplet and its composition during the nucleation process.
\end{abstract}

\maketitle

\section{Introduction}
The nucleation of liquid droplets is a ubiquitous phenomenon, central to many applications in chemistry, physics and atmospheric sciences~\cite{yasuoka1998molecular,oxtoby1992homogeneous,shen1999computational,weakliem1993toward,schenter1999dynamical,zeng1991gas,yi2002molecular,kinjo1999computer,toxvaerd2001molecular,ford1996thermodynamic,talanquer1995density,reiss1990molecular,kalikmanov1995semiphenomenological,horsch2008modification,neimark2005birth,oxtoby1988nonclassical,lutsko2008density,wang2008homogeneous,ten1999numerical,gonzalez2015bubble,loeffler2015improved,sosso2016crystal,xu2015effect,keasler2015understanding,wilhelmsen2015influence,van2015mechanism,hale1986application,hale2005temperature,hale2010scaled,yuhara2015nucleation,lauricella2015clathrate,singh2014characterization,ni2013effect,reinhardt2014effects,ten1998computer,chen2001aggregation,oh1999formation,chen2002simulating,zhukhovitskii1995molecular,nishi2015molecular,lupi2016pre,santiso2015general,berryman2016early,zimmermann2015nucleation,lam2015atomistic,kratzer2015two,bolhuis2015practical,lau2015water,toxvaerd2016nucleation}. Nucleation from a single component system can be rationalized in terms of a change in a single intrinsic variable, such as e.g. the chemical potential or pressure. In this case, the supersaturation of the parent vapor phase is simply characterized by a given value of the pressure~\cite{ten1998computer,tanaka2005tests,kraska2006molecular,oh2000small,senger1999molecular,lau2015water,toxvaerd2016nucleation} (or equivalently of the chemical potential) that departs from the pressure at coexistence. On the other hand, nucleation from a mixture involves a myriad of pathways arising from the larger dimension of the system, with as additional intrinsic variables, the mole fractions for each of the components~\cite{kulmala1990binary,zeng1991binary,oxtoby1994general,napari1999gas,jaecker1988nucleation,talanquer1995nucleation,ten1998numerical,laaksonen1995gas,yoo2001monte,napari2000surfactant,braun2014molecular,shimizu2015novel,pinho2014microfluidic,gao2014,alekseechkin2015thermodynamics,watson2011crystal,desgranges2014unraveling}. This also makes the definition of an appropriate reaction coordinate for the system especially challenging since, for instance, the total number of particles in the cluster is not the only significant variable, as the numbers of particles of each type also need to be taken into account to fully characterize the nucleation process. Here, we propose and implement a new approach that uses the entropy $S$, which captures the interplay between size increase and molecular selectivity during nucleation, as the reaction coordinate for the process.

The aim of this work is to shed light on the entropic pathways followed during the nucleation of two binary mixtures. The first example we consider is a binary mixture of two highly miscible gases, specifically the $Ar-Kr$ mixture. The second example involves a mixture of carbon dioxide with an alkane (here, as an example we consider $C_2H_6$) that is of technological relevance for the oil industry and for separation applications. We develop here the $\mu_1 \mu_2 VT-S$ method, where $\mu_1$ and $\mu_2$ are the chemical potentials for the two components of the mixture and $V$ and $T$ are the volume and temperature of the system, to simulate the nucleation of a liquid droplet for these two binary mixtures. In the first paper of the series~\cite{muVTS1}, we discussed how, in the case of a single component system, the $\mu VT-S$ approach provided a direct connection with classical nucleation theory~\cite{mcgraw1996scaling}, as any arbitrary value of $\mu$ (or, equivalently, of the supersaturation $\Delta \mu$) could be applied, and allowed the system to overcome the free energy barrier of nucleation. In practice, this was achieved by using the entropy of the system, which can be readily calculated in the grand-canonical ensemble, as the reaction coordinate and by driving the system along an entropic pathway using the umbrella sampling simulation technique. Extending this approach to the case of mixtures is especially appealing since the grand-canonical ensemble naturally allows the number of molecules of each component to vary as nucleation proceeds. It is therefore very well suited to shed light on the impact of the choice of a given supersaturation (i.e. $\Delta \mu_1$ and $\Delta \mu_2$ for a binary mixture) on the free energy barrier of nucleation as well as on the interplay between the size of the droplet and its composition as nucleation takes place.

The paper is organized as follows. In the next section, we present the simulation method as well as the molecular models used in this work. We explain how we set up the $\mu_1 \mu_2 V T-S$ simulations, detailing how we proceed with the choice of chemical potentials, supersaturations, range of entropies to be sampled and the relation between the conditions of nucleation and the composition of the bulk. We then discuss the results obtained during the simulations of the nucleation process from supersaturated vapor phases of $Ar-Kr$ and $CO_2-C_2H_6$. For all systems, we determine the free energy profile of nucleation and show how the choice of the conditions of nucleation impacts the height of the free energy barrier. We also focus on the analysis of the nucleation mechanism and on unraveling the interplay between size and composition during the formation of the liquid droplet, before finally drawing the main conclusions from this work in the last section.

\section{Simulation Method}

\subsection{Spanning entropic pathways}
We extend to the case of mixtures the simulation method developed in the first paper in this series~\cite{muVTS1}. The approach proposed is termed as $\mu VT-S$ and consists in sampling configurations of the system around a value of the entropy $S_0$ in the grand-canonical ($\mu VT$) ensemble. As discussed in the first part of this series, the $\mu VT-S$ simulation method provides a direct connexion with classical nucleation theory~\cite{mcgraw1996scaling}, since simulations of the nucleation process can be carried out for any value of the supersaturation $\Delta \mu$. Another advantage of this method is that it allows the calculation of the entropy of the system $S$. In the case of binary mixtures, we therefore define the method as $\mu_1 \mu_2 VT-S$ and evaluate the entropy during the simulations through

\begin{equation}  
S={{U-N_1\mu_1-N_2\mu_2} \over {T(N_1+N_2)}}
\end{equation}

where $\mu_1$ and $\mu_2$ are the chemical potentials for the binary mixture, $N_1$ and $N_2$ are the number of atoms/molecules for each of the two components and $U$ is the internal energy for the entire system given by, in the case of a binary mixture of atoms, 

\begin{equation}  
U=U_{pot}+{3 \over 2}N_1 k_BT+ {3 \over 2}N_2 k_BT
\end{equation}
where $U_{pot}$ is the potential energy for the system. In the case of linear molecules, we add a contribution of $k_BT$ per molecule to account for the rotational degrees of freedom.

The $\mu_1 \mu_2 VT-S$ simulation relies on gradually decreasing the entropy of the system through the application of a bias potential. This bias potential is defined within the framework of the umbrella sampling technique as 

\begin{equation}  
U_{bias}={1 \over 2} k (S-S_0)^2
\label{bias}
\end{equation}
in which $S_0$ is the target value for the entropy, $S$ is the current value of the entropy of the system and $k$ is a spring constant. This bias potential is then added to the potential energy of the system, and the total potential energy is used in the conventional Metropolis criteria for the acceptance of the different types of Monte-Carlo ($MC$) steps. For the $Ar-Kr$ mixture, $MC$ steps include the insertion ($12.5$~\% of the attempted $MC$ steps) or deletion ($12.5$~\% of the attempted $MC$ steps)  of atoms as well as the translation of a single atom ($75$~\% of the attempted $MC$ steps). In the case of the $C_2H_6-CO_2$ mixture, we have the following rates: rotation ($37.5$~\% of the attempted $MC$ steps), translation ($37.5$~\%), insertion ($12.5$~\%) and deletion ($12.5$~\%). 

Successive umbrella simulations with decreasing values for the target entropy ($S_0$) are carried out to achieve the formation of a liquid droplet of a critical size. During each of these simulations, histograms for the number of times a given entropy interval is visited are collected, allowing for the calculation of the free energy profile associated with the nucleation process~\cite{torrie1977nonphysical,Allen,Au,Alu2,desgranges2014unraveling}.

\subsection{Simulation models}
The simulation models used in this work for $Ar$ and $Kr$ are based on the Lennard-Jones potential~\cite{vrabec2001set}, with the following parameters for $Ar$: $\sigma_{Ar}$=$3.3952$~\AA and $\epsilon_{Ar}/k_B$=$116.79~K$ and for $Kr$: $ \sigma_{Kr}$=$3.6274$~\AA and $\epsilon_{Kr}/k_B$=$162.58$~K.

For the $CO_2-C_2H_6$, we use a force field that models the dispersion-repulsion interactions through an exp-6 functional form~\cite{Potoff,Errington1,Errington2,Errington3}. 

\begin{equation}
\begin{array}{lll}
u(r) &= {\epsilon \over {1-6/\alpha}} \left[ {6 \over \alpha} exp \left( \alpha \left[ 1 - {r \over r_m} \right] \right) - \left( {r_m \over r}  \right)^6  \right] & (r>r_{max}) \\
& = \infty & (r<r_{max}) \\
\label{exp6}   
\end{array}
\end{equation}

In Eq.~\ref{exp6}, $r_m$ is the distance for which the potential reaches a minimum, $r_{max}$ is the smallest positive distance for which $du(r)/dr=0$ and $\epsilon$ and $\alpha$ are two potential parameters. As discussed by Errington {\it et al.}~\cite{Errington1}, it is convenient to discuss the potential parameters in terms of $\sigma$ (i.e. the distance for which $u(r)=0$, obtained numerically by solving the equation $u(\sigma)=0$) rather than in terms of $r_m$. In the case of $C_2H_6$, we use an united atom-type force field and model the molecule with two $\exp-6$ sites, each site standing for a $CH_3$ group. We use the following set of parameters: $\epsilon_{CH_3}/k_B= 129.64$~K, $\sigma_{CH_3}=3.679$~\AA~and $\alpha_{CH_3}=16$. In the case of $CO_2$, in addition to the the repulsion-dispersion interactions, a Coloumbic term is added to account for the quadrupolar nature of $CO_2$. The $CO_2$ molecule is modeled with a distribution of three $\exp-6$ sites and three point charges located on each of the atoms. We use the following parameters for the $\exp-6$ sites: $\epsilon_{C}/k_B= 29.07$~K, $\sigma_{C}=2.753$~\AA~and $\alpha_{C}=14$, $\epsilon_{O}/k_B= 83.20$~K, $\sigma_{O}=3.029$~\AA~ and $\alpha_{O}=14$. We also have $q_C=0.6466$~e and $q_O=-03233$~e. Both molecules are considered to be rigid, with a distance between the two $CH_3$ $\exp-6$ sites set to $1.839$~\AA~for ethane and a length of the $C-O$ bond fixed to $1.1433$~\AA~for $CO_2$. The calculation of the interaction energy is performed for distances up to $13.5$~\AA~for both systems with, electrostatic interactions calculated with the Ewald sum method beyond that distance using the parameters given in previous work~\cite{PartIII}. In line with previous simulation work on nucleation~\cite{ten1998computer}, we do not include any tail corrections for the dispersion-repulsion interactions beyond the cutoff distance. 

\subsection{Setting up the simulation}

\subsubsection{$Ar-Kr$ mixture}

We carry out $\mu_1 \mu_2 VT-S$ simulations for the $Ar-Kr$ mixture at $T=148.15$~K in cubic cells with an edge of $100$~\AA~ (the usual periodic boundary conditions are applied). We take advantage of the Expanded Wang-Landau (EWL) method we recently developed~\cite{PartI,PartII,PartIII,PartIV} to obtain very accurate estimates for $\mu_1$ and $\mu_2$ both at the vapor-liquid coexistence and for supersaturated vapor phases, which will be the parent phase for the nucleation events. The values are presented in Table~\ref{muAr}

\begin{table}[hbpt]
\caption{$Ar-Kr$ mixture at $148.15$~K: chemical potentials for $Ar$ $(\mu_1)$ and $Kr$ $(\mu_2)$, supersaturations, mole fractions in $Kr$ for the vapor ($y_{Kr}$) and for the liquid ($x_{Kr}$), pressure and entropies for the two coexistence points (coex I and coex II) and for the supersaturated vapors considered in this work (system~1 to 4).}
\begin{tabular}{|c|c|c|c|c|c|c|c|c|c|c|}
\hline
$ $ & $\mu_1$ & $\mu_2$ & $\Delta \mu_1$ & $\Delta \mu_2$ & $x_{Kr}$ & $y_{Kr}$ & $P$ & $P/P_{coex}$ & $S_l$ & $S_v$\\
$ $ & $(kJ/mol)$ & $(kJ/mol)$ & $(kJ/mol)$ & $(kJ/mol)$ & $ $ & $ $ & $bar$ & $ $ & $(kJ/mol/K)$ & $(kJ/mol/K)$\\
\hline
\hline
 coex~I& -14.077 & -17.571 & - & - & 0.440 & 0.190 & 28.68 & 1.0 & 0.0848 & 0.1074  \\
\hline
 system~1 & -13.951 & -17.463 & 0.049 & 0.108 & 0.440 & - & 57.27 & 2.0 & 0.0831 & -  \\
\hline
 system~2 & -13.930 & -17.439 & 0.070 & 0.132 & 0.440 & - & 63.09 & 2.2 &  0.0828 & - \\
\hline
\hline
 coex~II & -13.720 & -18.165 & - & - & 0.250 & 0.108 & 36.92 & 1.0 & 0.0845 & 0.1021  \\
\hline
 system~3 & -13.548 & -18.057 & 0.172 & 0.108 & 0.250 & - & 73.83 & 2.0 &  0.0817 & - \\
\hline
 system~4 & -13.514 & -18.039 & 0.206 & 0.126 & 0.250 & - & 81.28 & 2.2 & 0.0813 & -  \\
\hline
\hline
\end{tabular}
\label{muAr}
\end{table}

We start by determining the conditions for the vapor-liquid coexistence at the coex I point. In practice, this is done using the EWL method by finding numerically the values for $\mu_1$ and $\mu_2$ which lead to equal probabilities for the vapor and the liquid phase (see more details in previous work~\cite{PartIII}). The EWL method also allows to obtain all thermodynamic properties for the mixture including the mole fractions, pressure as well as the entropies for the two coexisting phases. These entropies, which are also given in Table~\ref{muAr}, provide an idea of the range of entropies that need to be sampled for the system to undergo the vapor $\to$ liquid transition. From the coexistence point, we can increase the value of the two chemical potentials $\mu_1$ and $\mu_2$ or, in other words, create a supersaturated vapor that will serve as a starting point for the nucleation process. We list in Table~\ref{muAr} the two sets of supersaturations (system 1 and system 2) generated from the coexistence point coex I. As can be seen from Table~\ref{muAr}, increasing $\Delta \mu$ brings the supersaturated vapor more deeply into the liquid domain of the phase diagram, resulting in a larger value for the pressure and a lower value for the entropy of the liquid. We proceed along the same lines from the second coexistence point (coex II) to define two supersaturated vapors (system 3 and system 4), this time with a mole fraction in the liquid ($x_{Kr}$ set to $0.250$). We finally add that there are other ways of determining the chemical potential at the vapor-liquid coexistence and for supersaturated vapors~\cite{Camp,Pana,Potoff,nezbeda1991new,Singh,rai2007pressure,Rane,expanded,Shi1,eslami2007molecular,Vogt,widom1963some,CBMC}.

\begin{figure}
\begin{center}
\includegraphics*[width=8cm]{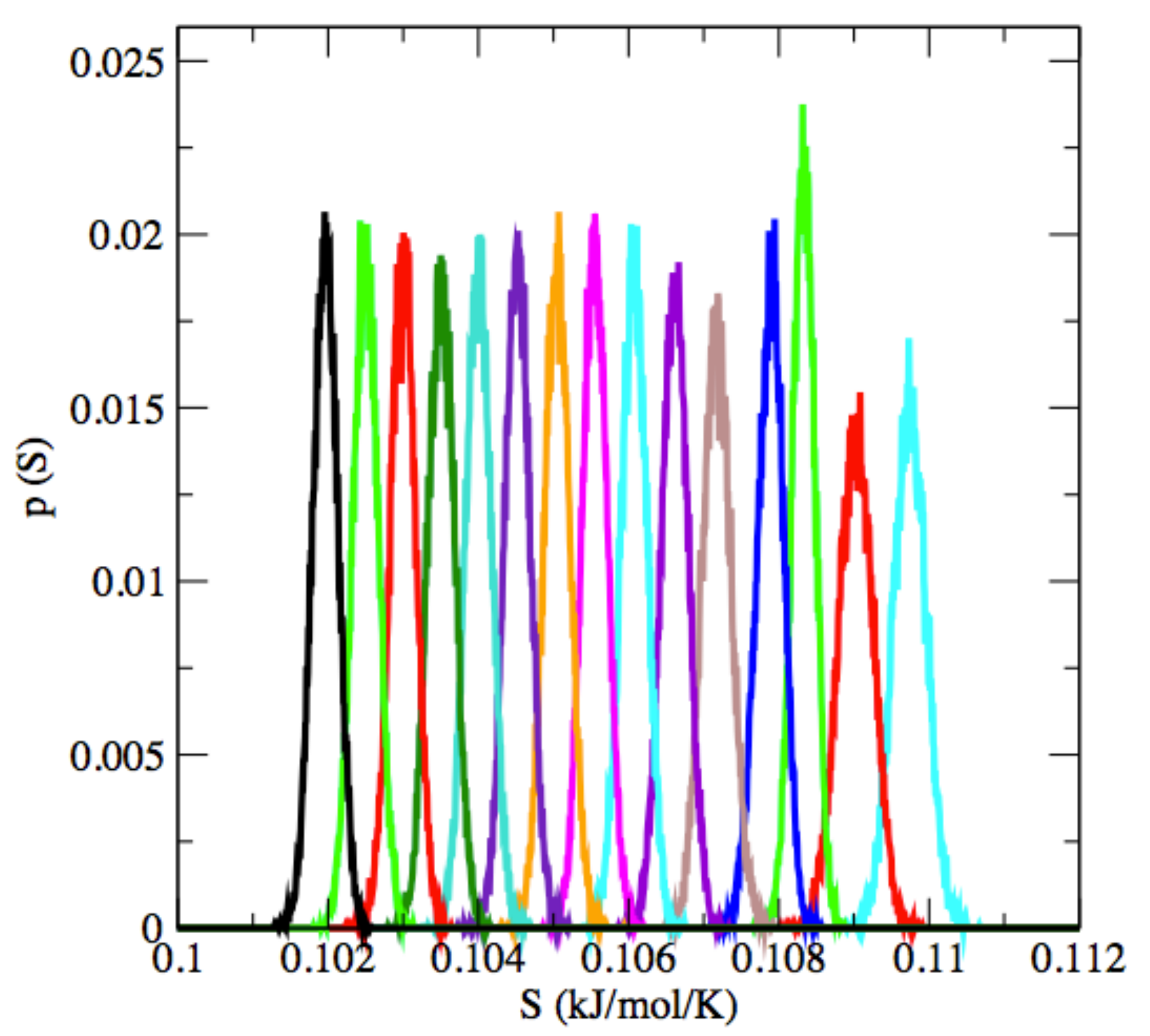}
\end{center}
\caption{$Ar-Kr$ mixture (system 1). Histograms $p(S)$ collected during the umbrella sampling simulations for decreasing values for the entropy.}
\label{Fig1}
\end{figure}

We plot in Fig.~\ref{Fig1} the successive umbrella sampling windows carried out during the $\mu_1 \mu_2 VT-S$ simulations on system 1. Fig.~\ref{Fig1} shows the histograms corresponding to the probability according to which a given entropy interval is visited during the $\mu_1 \mu_2 VT-S$ simulations. Each of the umbrella sampling windows (labeled with an index $i$) is obtained by imposing through Eq~\ref{bias} a different value of the target entropy $S_{0,i}$. During the nucleation of a liquid droplet, the system goes from a supersaturated vapor with a low density, and thus of high entropy, to a system containing a liquid droplet of a critical size, i.e. to a much more dense system of lower entropy. To observe the formation of the liquid droplet, we therefore carry out successive umbrella sampling windows for decreasing values of the target entropy $S_{0,i}$ and obtain the histograms, shown in Fig.~\ref{Fig1}, that cover the entire nucleation process. The progress of the system towards the formation of a liquid droplet can be followed by monitoring the relative location of $S_{max,i}$, the entropy for which the histogram $p_i(S)$ reaches its maximum, and of $S_{0,i}$, the target entropy for the $i^{th}$ umbrella sampling window. We start with the window located to the right of Fig.~\ref{Fig1}, associated with the largest value of $S_{0,1}$ which corresponds to a very dilute vapor ($S_{0,1}=0.11$~kJ/mol/K). As shown in Fig.~\ref{Fig1}, for the first window (starting from the right), we have $S_{max,1}=0.1097$~kJ/kg/K, which is less than the target value $S_{0,1}=0.11$~kJ/kg/K. This means that the target value $S_{0,1}$ is greater than the entropy of the metastable supersaturated vapor for the choice of $(\mu_1,\mu_2)$ made for system 1. Gradually decreasing the target entropy for the next windows allows us to find the value $S_{0,i}$ which coincides with the maximum for $p_i(S)$. This occurs here for $S_{max,i}=S_{0,i}=0.1091$~kJ/kg/K. At this point, we have the metastable supersaturated vapor. Then, during the next few umbrella sampling windows, we observe a change in behavior as the histograms $p_i(S)$ now lag behind the target value for the entropy with $S_{max,i}>S_{0,i}$. This corresponds to the fact that the system has to overcome the free energy cost in forming the liquid droplet. Later on, for $S_{0,i}=0.103$~kJ/kg/K, we find again that the maximum for $p_i(S)$ coincides with $S_{0,i}$, indicating that we have reached the top of the free energy barrier of nucleation and that a liquid droplet of a critical size has formed. For target values of the entropy greater than $0.103$~kJ/kg/K, we observe again a change in behavior as the histograms $p_i(S)$ run ahead of the target value for the entropy with $S_{max,i}<S_{0,i}$, corresponding to the spontaneous growth of the liquid droplet.

\begin{figure}
\begin{center}
\includegraphics*[width=8cm]{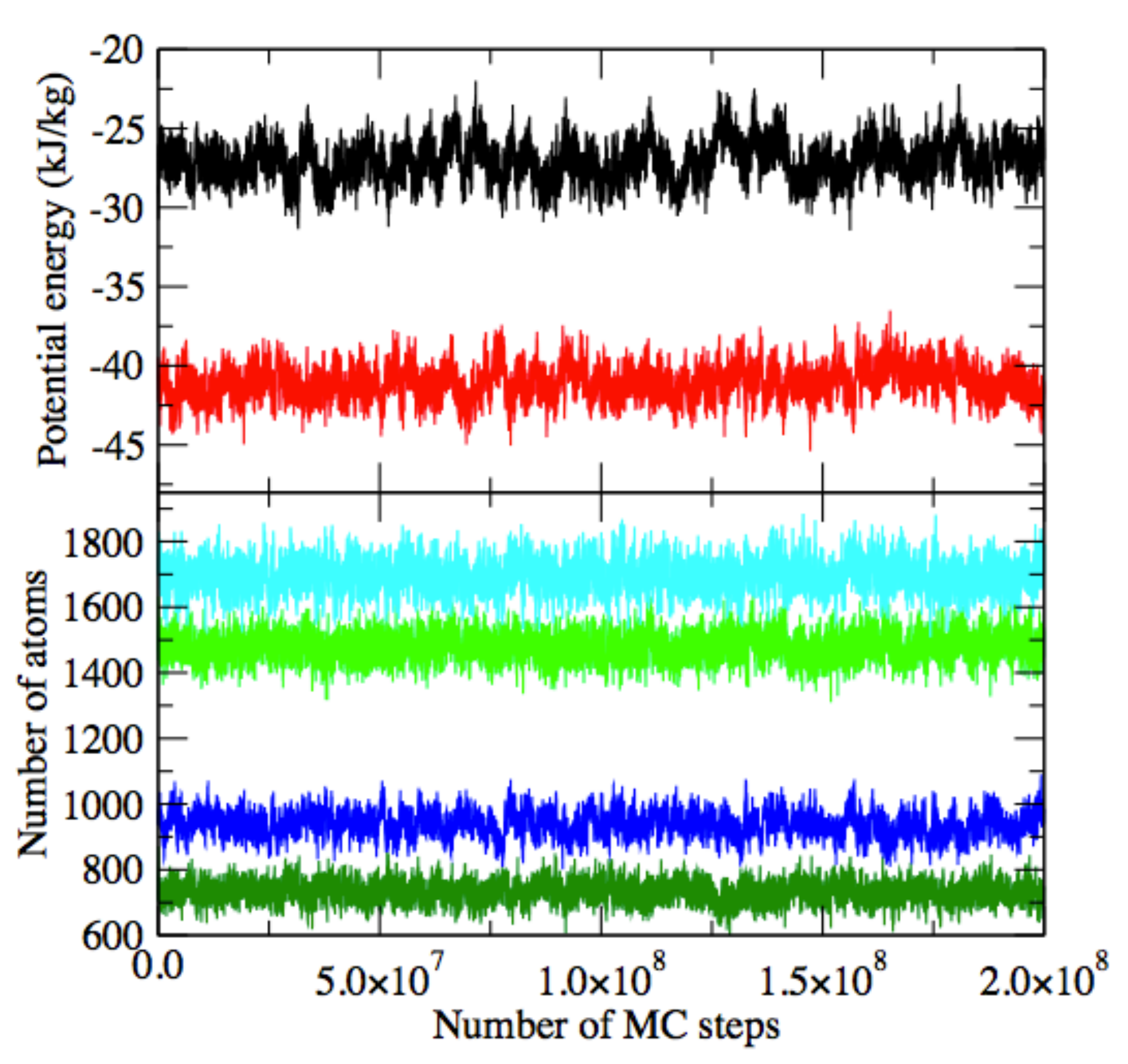}
\end{center}
\caption{$Ar-Kr$ mixture for a supersaturation of $2$ at $x_{Ar}=0.56$ and at $T=148.15$~K. (Top) Potential energy of the system during a simulation at entropy fixed to $S_0=0.107$~kJ/mol/K (black) and $S_0=0.104$~kJ/mol/K (red). (Bottom) Number of atoms during the simulation at $S_0=0.107$~kJ/mol/K ($N_{Ar}$: light green and $N_{Kr}$: dark green) and at $S_0=0.104$~kJ/mol/K ($N_{Ar}$: cyan and $N_{Kr}$: blue).}
\label{Fig2}
\end{figure}

To provide additional insight in the evolution of the average properties of the system for different umbrella sampling windows, we show in Fig~\ref{Fig2} the potential energy and the number of atoms for different values of the target entropy $S_{0,i}=0.107$~kJ/mol/K and $S_{0,i}=0.104$~kJ/mol/K during a production run. These plots show the impact of decreasing the target value for the entropy on the system. For example, comparing the results for two different umbrella sampling windows, $0.107$~kJ/mol/K and $0.104$~kJ/mol/K, we observe a decrease in the potential energy by about $60$~\%. This happens simultaneously with an increase in the number of atoms of each component in the mixture by about $15$~\% for $Ar$ and by $20$~\% for $Kr$. This confirms that the decrease in the target value for the entropy occurs with a greater organization and density of the system, that we will analyze in depth in the 'Results' section. We finally add that, for each umbrella sampling window, we run an equilibration run of $100\times10^6$ MC steps, followed by a production run of $200\times10^6$ MC steps. Throughout the $\mu_1 \mu_2 VT-S$ simulations, we also check that the acceptance rate for the insertion/deletion steps remains high enough to ensure an accurate sampling of the configurations of the systems. For instance, in the case of system 1 and for the configurations of highest density (umbrella sampling window with a critical liquid droplet for $S_{0,i}=0.103$~kJ/mol/K), the acceptance rates for the insertion/deletion steps are of $45.8$~\%. 

\subsubsection{$C_2H_6-CO_2$ mixture}

Simulations of droplet nucleation for the $C_2H_6-CO_2$ mixture are performed at $T=263.15$~K in cubic cells, with an edge of $100$~\AA, and with the usual periodic boundary conditions. We use EWL simulations to determine the other input parameters for the $\mu_1 \mu_2 VT-S$ simulations and list in Table~\ref{muCO2} the sets of chemical potentials $(\mu_1,\mu_2)$ used for these simulations.

\begin{table}[hbpt]
\caption{$C_2H_6-CO_2$ mixture at $263.15$~K: chemical potentials for $C_2H_6$ $(\mu_1)$ and $CO_2$ $(\mu_2)$, supersaturations, mole fractions in $CO_2$ for the vapor ($y_{CO_2}$) and for the liquid ($x_{CO_2}$), pressure and entropies for the two coexistence points (Coex~III and Coex~IV) and for the supersaturated vapors considered in this work (system~5 to 8).}
\begin{tabular}{|c|c|c|c|c|c|c|c|c|c|c|}
\hline
$ $ & $\mu_1$ & $\mu_2$ & $\Delta \mu_1$ & $\Delta \mu_2$ & $x_{CO_2}$ & $y_{CO_2}$ & $P$ & $P/P_{coex}$ & $S_l$ & $S_v$\\
$ $ & $(kJ/mol)$ & $(kJ/mol)$ & $(kJ/mol)$ & $(kJ/mol)$ & $ $ & $ $ & $bar$ & $ $ & $(kJ/mol/K)$ & $(kJ/mol/K)$\\
\hline
\hline
 coex~III & -38.186 & -44.397 & - & - & 0.053 & 0.124 & 27.37 & 1.0 & 0.1397 & 0.1556  \\
\hline
 system~5 & -38.055 & -44.268 & 0.131 & 0.129 & 0.053 & - & 43.80 & 1.6 &  0.1318 & - \\
\hline
 system~6 & -38.036 & -44.250 & 0.150 & 0.147 & 0.053 & - & 46.53 & 1.7 & 0.1316 & -  \\
\hline
\hline
 coex~IV& -38.261 & -43.216 & - & - & 0.097 & 0.201 & 29.60 & 1.0 & 0.1409 & 0.1566  \\
\hline
 system~7 & -38.121 & -43.076 & 0.140 & 0.140 & 0.097 & - & 47.35 & 1.6 & 0.1333 & -  \\
\hline
 system~8 & -38.099 & -43.061 & 0.162 & 0.155 & 0.097 & - & 50.32 & 1.7 &  0.1331 & - \\
\hline
\hline
\end{tabular}
\label{muCO2}
\end{table}

We identify a first state point, coex III, leading to vapor-liquid coexistence for the $C_2H_6-CO_2$ mixture. As previously discussed, this is achieved by finding numerically $\mu_1$ and $\mu_2$ such that the liquid and the vapor phases are equally probable. The EWL simulations also yield the mole fractions as well as the entropies of the two coexisting phases, bracketing the range of entropies needed to sample the nucleation of the liquid droplet. From coex III, we increase the chemical potentials of the two mixture components by $\Delta \mu_1$ and $\Delta \mu_2$ to obtain thermodynamic conditions located in the domain of the liquid in the phase diagram and, for which, we can observe a supersatured vapor. We repeat this step for two different supersaturations leading to system 5 and 6 in Table~\ref{muCO2} (as for $Ar-Kr$, the choices for $\Delta \mu_1$ and $\Delta \mu_2$ are made such that the liquid mole fraction in the second component, here $CO_2$, remains constant). We choose another coexistence point, coex IV, and two corresponding supersaturated vapors, system 7 and system 8. For each umbrella sampling window, an equilibration run of $50\times10^6$ MC steps is run, followed by a production run of $100\times10^6$ MC steps. As with the $Ar-Kr$ mixture, we check that the acceptance rates for the insertion/deletion steps remains high enough to ensure an accurate sampling. This is the case here for all umbrella sampling windows. For instance, in the case of system 5 and for the window associated with the highest density (umbrella sampling window with a critical liquid droplet for $S_{0,i}=0.155$~kJ/mol/K), the acceptance rates for the insertion/deletion steps are of $36.9$~\% for $C_2H_6$ and of $46.6$~\% for $CO_2$.

\section{Results and Discussion}

\subsection{$Ar-Kr$ mixture}

\begin{figure}
\begin{center}
\includegraphics*[width=8cm]{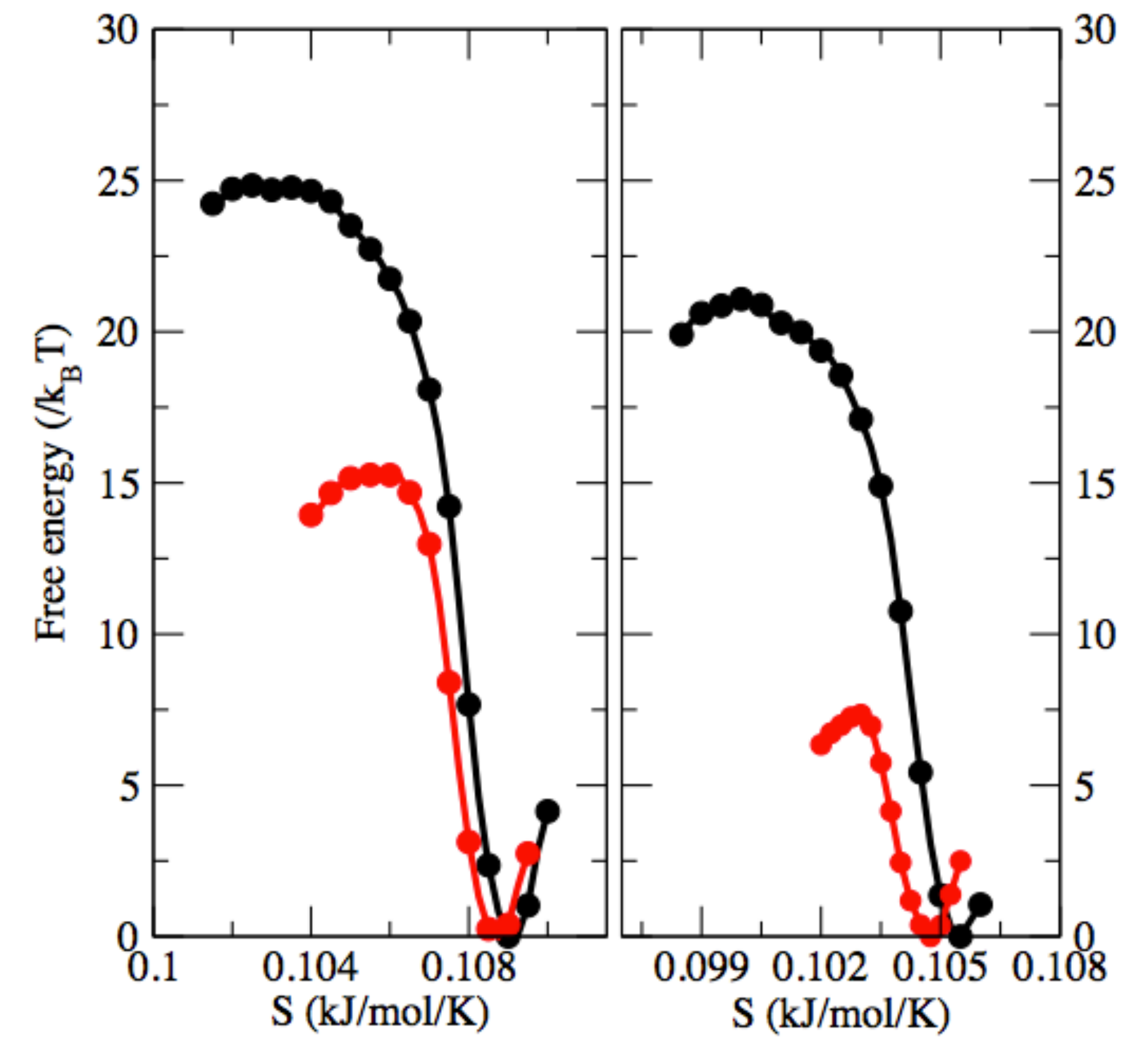}
\end{center}
\caption{Free energy barriers of nucleation for the $Ar-Kr$ mixture at $T=148.15$~K. (Left panel) system 1 (black) and system 2 (red) for $x_{Ar}=0.56$. (Right panel) system 3 (black) and system 4 (red) for $x_{Ar}=0.75$}
\label{Fig3}
\end{figure}

We start by analyzing the free energy barriers obtained for the $Ar-Kr$ mixture at $T=148.15$~K. The left panel of Fig.~\ref{Fig3} shows the free energy profiles for two different supersaturations, system 1 and system 2, at $x_{Kr}=0.44$. The supersaturation has a direct effect on the height of the free energy barrier, with a barrier of $25\pm2~k_BT$ for system 1 and of $15\pm2~k_BT$ for a system 2. $\Delta \mu_1$ and $\Delta \mu_2$ are greater for system 2, which means that the parent supersaturated vapor of system 2 is located more deeply into the domain of the liquid in the phase diagram, and, as such, that the nucleation of a liquid droplet occurs more easily. This also leads to a lower free energy of nucleation than for system 1. We find that the range of entropies spanned during the nucleation process is also impacted by the amount of supersaturation. For instance, looking at system 2 in Fig.~\ref{Fig3}, we find that, for the higher supersaturation, the passage from the supersaturated vapor to a system containing a droplet of a critical size occurs over entropies between $0.1085$~kJ/mol/K, for which the free energy reaches a minimum for the entropy of the metastable parent phase, and $0.1055$~kJ/mol/K, for which the free energy reaches a maximum corresponding to the formation of a droplet of a critical size. On the other hand, for system 1 (lower supersaturation), we find that liquid nucleation takes place over a much broader entropy range, with the entropic pathway ranging from $0.1091$~kJ/mol/K to $0.103$~kJ/mol/K. This can be attributed to the combination of two effects. For a higher supersaturation, the density for the parent supersaturated vapor is larger and, therefore, its entropy is lower. Furthermore, at high supersaturation, the thermodynamic conditions lie further inside the liquid domain. Thus, the critical size for the liquid droplet becomes smaller, resulting in a system at the top of the free energy barrier that has a higher entropy. Both effects account for the narrowing, at high supersaturation, of the entropy range spanned during the droplet nucleation along the entropic pathway.

We now turn to the second set of simulations carried out for system 3 and system 4 at $x_{Kr}=0.25$. The free energy profiles obtained for the two supersaturations are shown in the right panel of Fig.~\ref{Fig3}. The behavior observed for $x_{Kr}=0.25$ exhibits similar qualitative features as for $x_{Kr}=0.44$. We find that the height of the free energy barrier of nucleation increases as we go from system 4 to system 3 (as the supersaturation decreases), with the free energy of nucleation increasing from $7\pm1~k_BT$ to $21\pm2~k_BT$. The range of entropies spanned during nucleation is also found to increase as the supersaturation decreases. For system 4, we observe that the entropy of the parent phase is of $0.1047$~kJ/mol/K while the entropy at the top of the free energy barrier is of $0.103$~kJ/mol/K. For a lower supersaturation (system 3), the entropy of the supersaturated vapor is $0.1055$~kJ/mol/K while it is of $0.100$~kJ/mol/K at the top of the free energy barrier. This means that a significantly broader range of entropies is sampled during the nucleation process. There are, however, notable differences between the two plots, which result from the interplay between pressure and the chemical composition of the system. Considering the results at fixed supersaturation, we find that the free energy of nucleation decreases by $16$~\% (going from system 1 to system 3), while it decreases by $53$~\% between system 2 and system 4. Similarly, the dependence of the height of the free energy barrier upon supersaturation is also shown to be impacted, with a decrease by $40$~\% from system 1 to system 2 at $x_{Ar}=0.56$ and a much larger decrease by $66$~\% from system 3 to system 4 at $x_{Ar}=0.75$. Both findings can be attributed to the fact that liquid mixtures with a lower fraction of $Kr$ are obtained for higher pressures. This, in turn, implies that the corresponding parent phases will be supersaturated vapors of larger densities and, therefore, result in lower free energy barriers of nucleation.
 
\begin{figure}
\begin{center}
\includegraphics*[width=5cm]{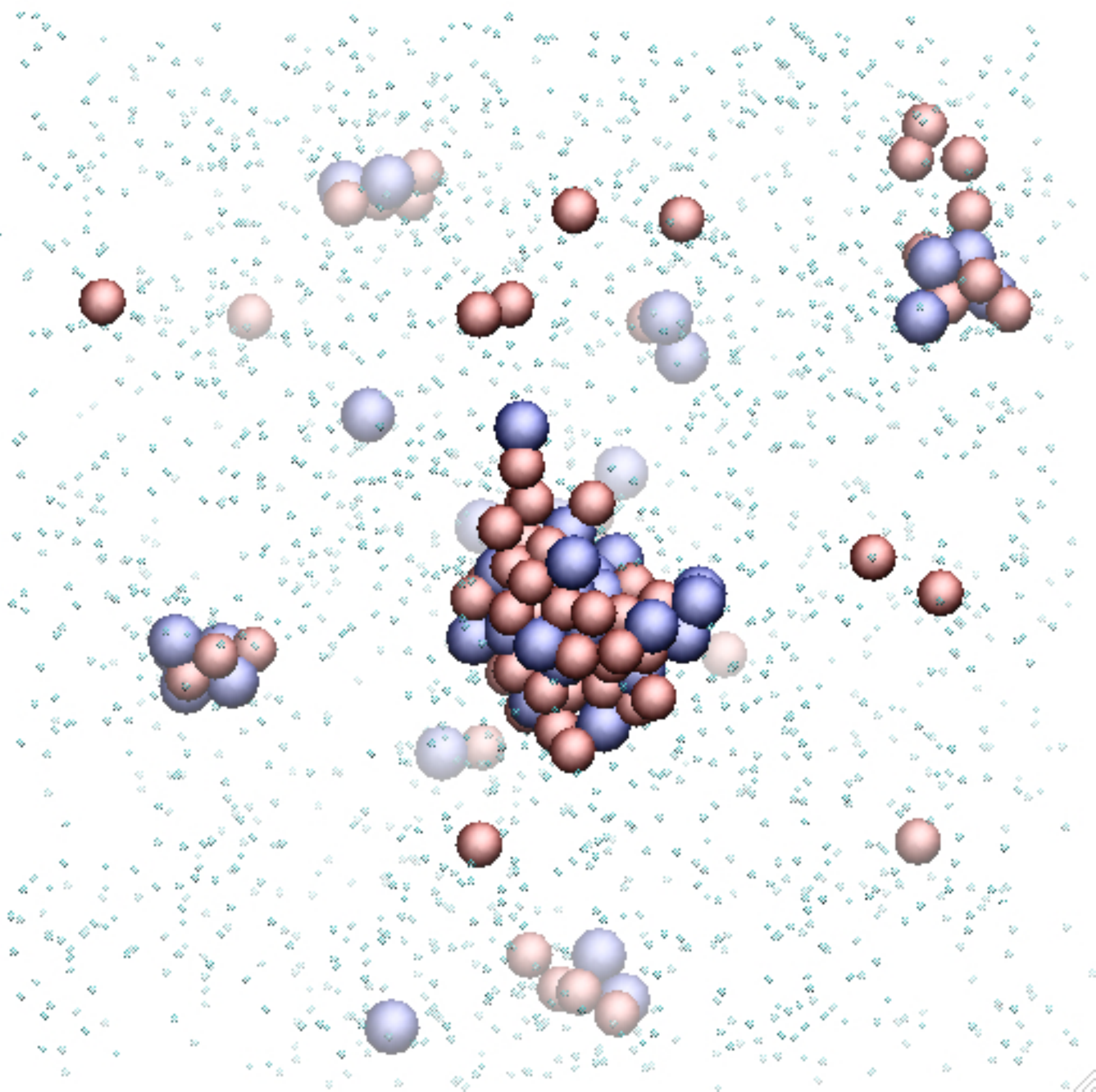}
\includegraphics*[width=5cm]{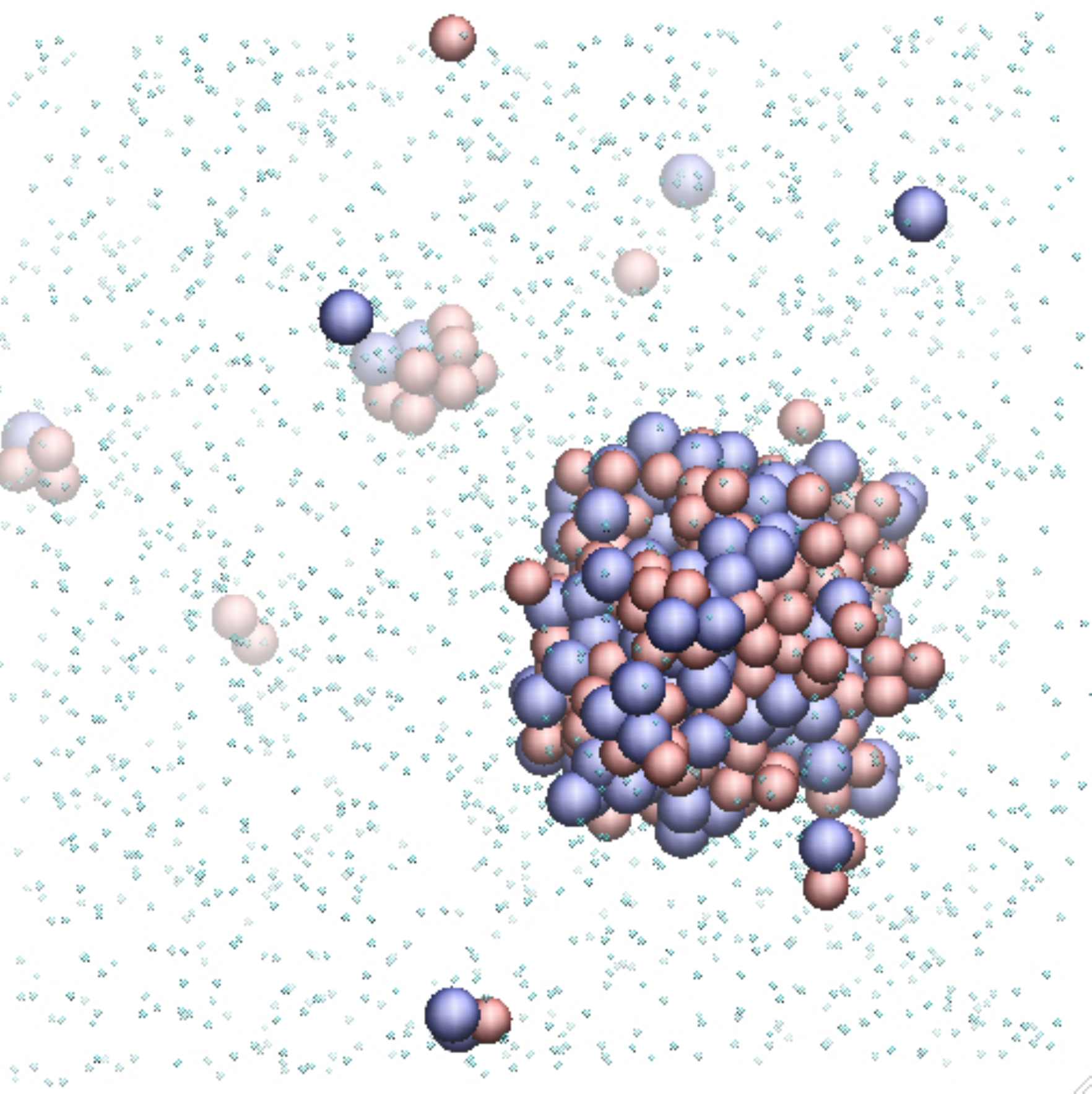}
\includegraphics*[width=5cm]{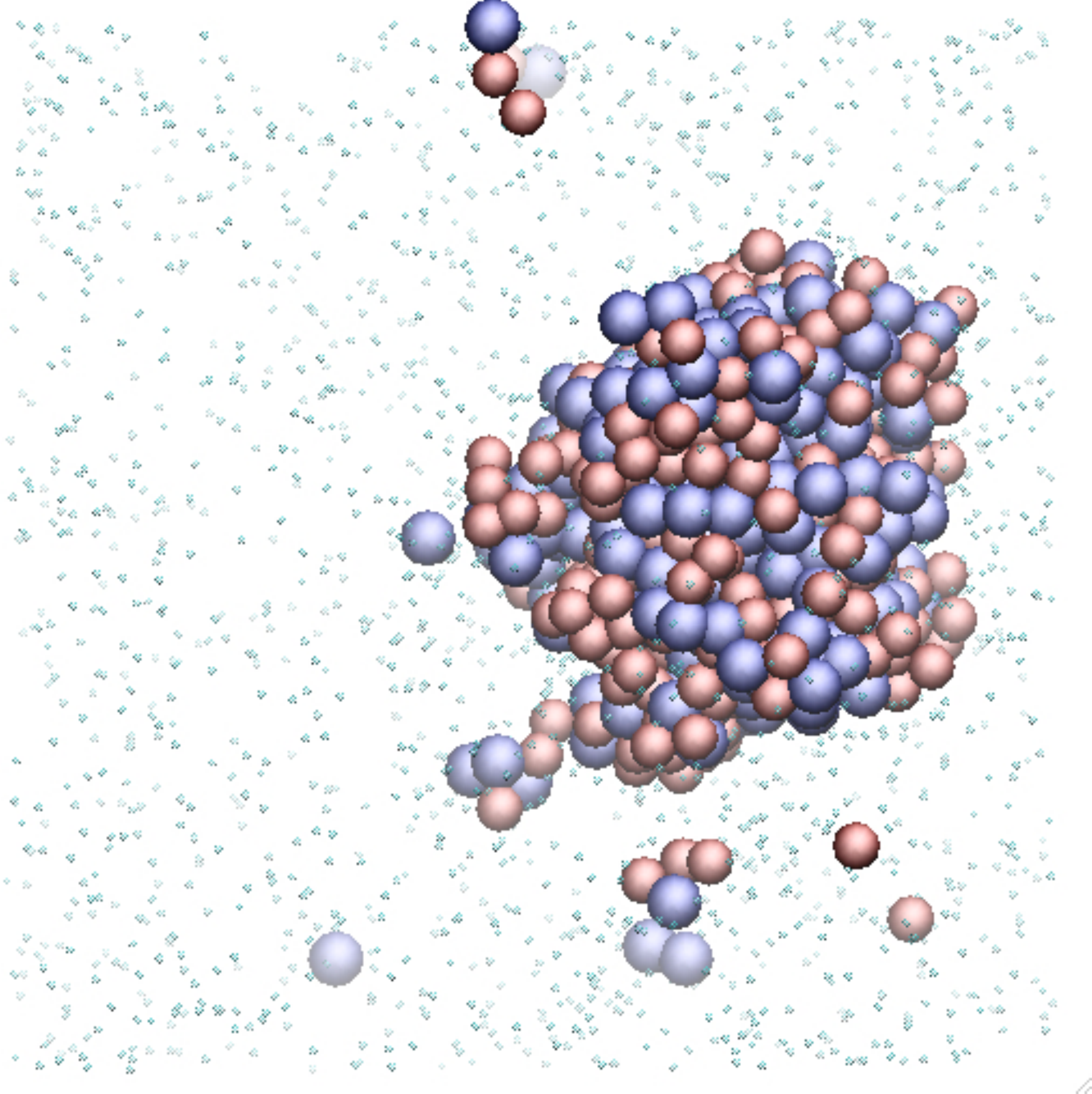}
\end{center}
\caption{$Ar-Kr$ mixture: Snapshots of system 1 during the nucleation process for a target entropy of $0.1075$~kJ/mol/K (left), $0.105$~kJ/mol/K (middle) and $0.103$~kJ/mol/K (right). Atoms with a liquid-like environment are shown with larger spheres, with $Ar$ in red and $Kr$ in blue. }
\label{Fig4}
\end{figure}

\begin{figure}
\begin{center}
\includegraphics*[width=8cm]{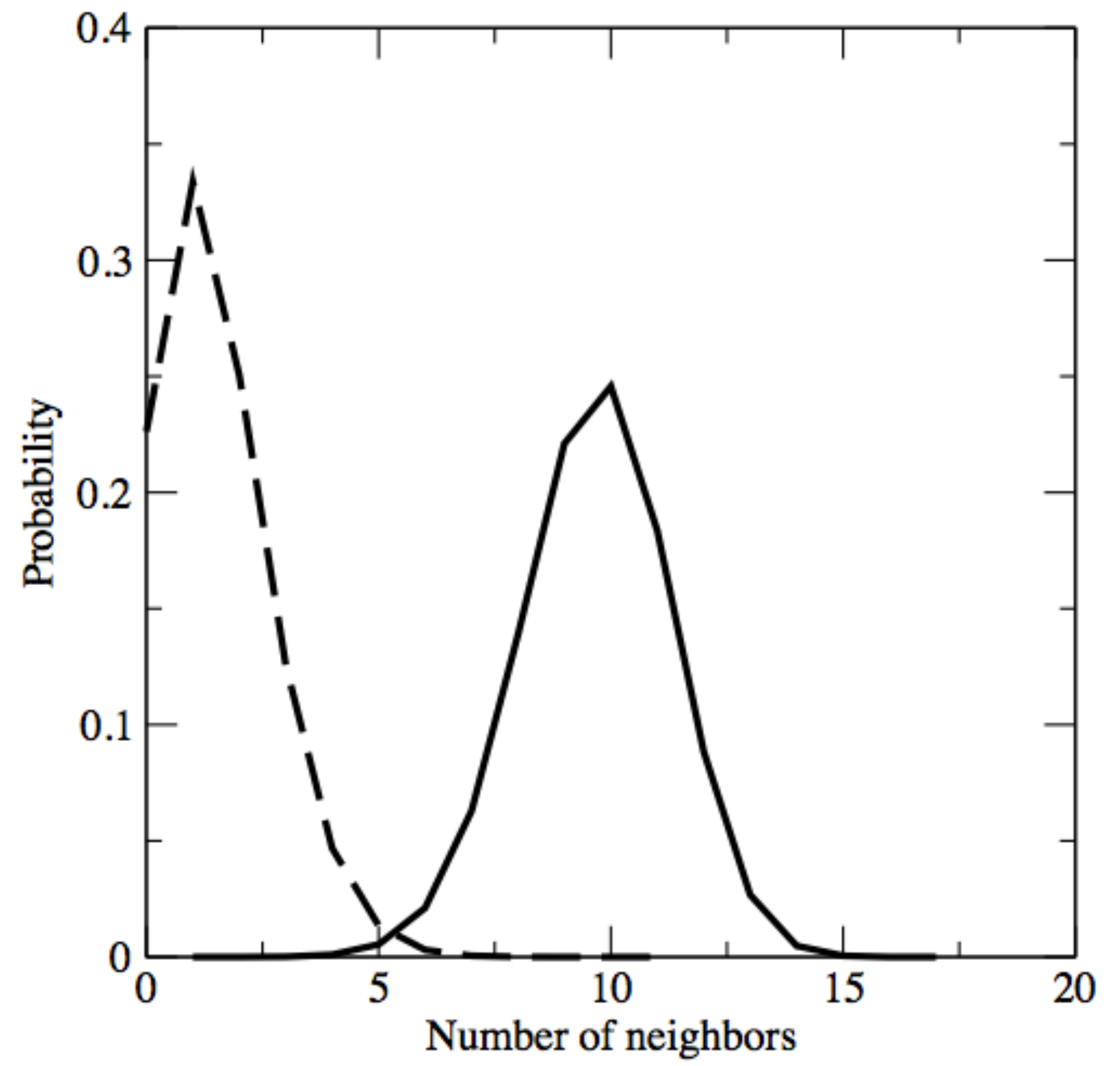}
\end{center}
\caption{$Ar-Kr$ mixture: Distributions for the number of neighbors within a distance of $5.4$~\AA~for the liquid (solid line) and for the vapor (dashed line) at coexistence for $T=148.15$~K.}
\label{Fig5}
\end{figure}

The plot shown in Fig.~\ref{Fig3} for the free energy profile of nucleation is a projection of the multi-dimensional free energy surface, as discussed in prior work on the nucleation in binary systems~\cite{reiss1950kinetics}. Here, we do not take into account any nonisothermal effect during nucleation~\cite{wyslouzil1992nonisothermal,wedekind2007influence}. To interpret further the results obtained for the free energy of nucleation as a function of S, we now focus on the interdependence between the entropy S, the droplet size and its composition. As we have seen from Fig.~\ref{Fig2}, the number of atoms steadily increases as the target entropy $S_{0,i}$ is decreased. To check that the decrease in entropy undergone by the system leads to an increased organization within the system and to the formation of a liquid droplet, we show in Fig.~\ref{Fig4} snapshots of the system, obtained during umbrella sampling simulations for decreasing values for the target entropy. As $S_{0,i}$ decreases, the droplet size increases, with a larger number of atoms being incorporated to the droplet as the target entropy decreases from $0.1075$~kJ/mol/K, to $0.105$~kJ/mol/K, and finally to $0.103$~kJ/mol/K. To assess further this point, we perform a detailed analysis of the size and composition of the droplet along the entropic pathway. The atoms belonging to the incipient droplet are identified through a commonly used geometric criterion~\cite{ten1998computer}. For this purpose, we determine the distribution for the number of neighbors within a distance of $5.4$~\AA~of a central atom in the vapor and the liquid mixture. We show in Fig.~\ref{Fig5} these distributions. Both distributions are sharply peaked around $N_{nab}=1$ for the vapor and around $N_{nab}=10$ for the liquid. As shown in Fig.~\ref{Fig5}, atoms belonging to the vapor always have less than $6$ neighbors within a distance of $5.4$~\AA. This allows us to introduce the following condition to identify a liquid-like atom, or equivalently an atom belonging to the developing droplet, as an atom with at least $6$ nearest neighbors within $5.4$~\AA.
 
\begin{figure}
\begin{center}
\includegraphics*[width=10cm]{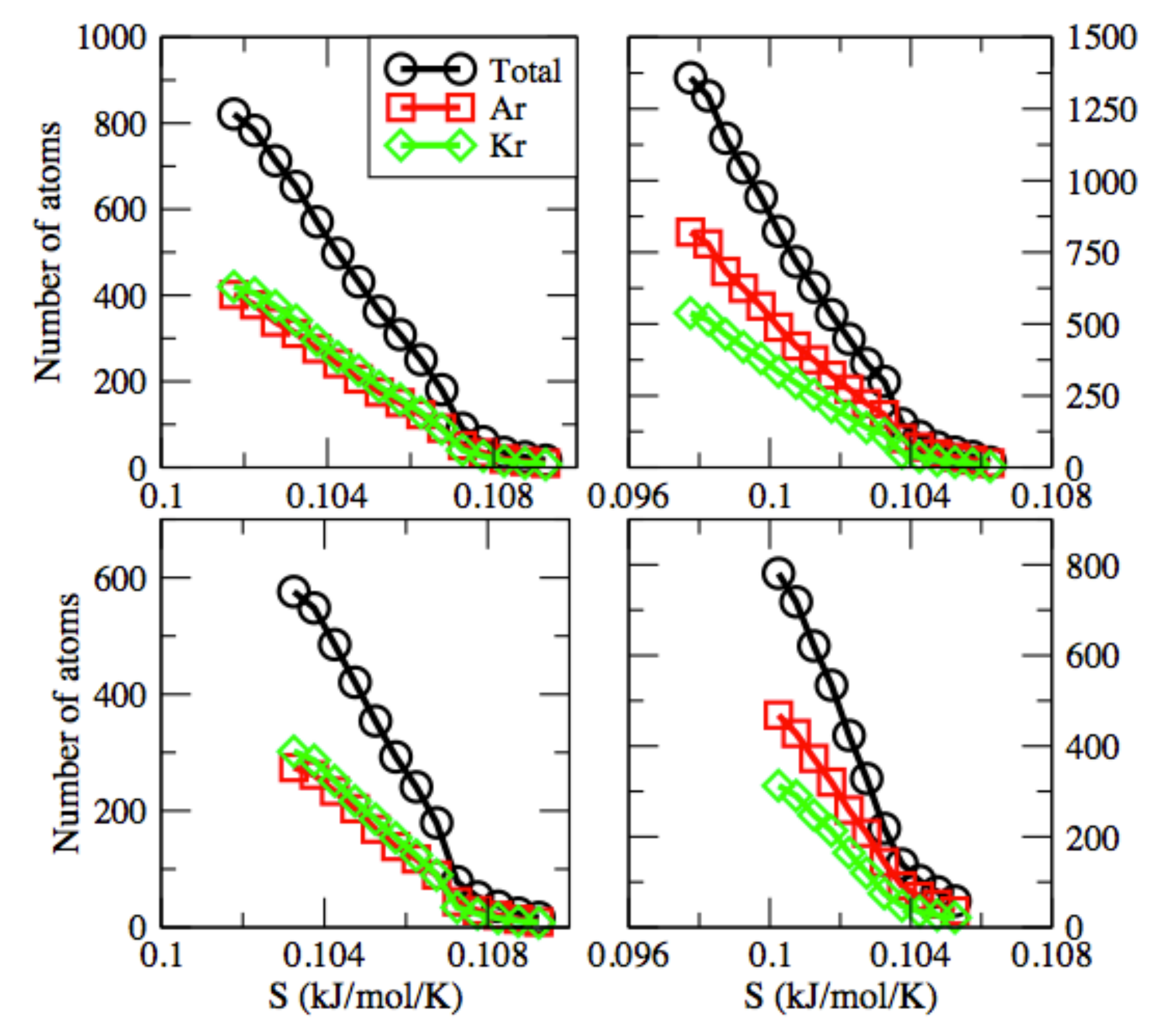}
\end{center}
\caption{$Ar-Kr$ mixture. (Left panel) Variation of the total number of atoms in the cluster and of the number of atoms for $Ar$ and $Kr$ at $x_{Ar}=0.56$ for system 1 (top) and system 2 (bottom). (Right panel) Total number of atoms in the cluster, number of atoms for $Ar$ and $Kr$ against $S$ at $x_{Ar}=0.75$ for system 3 (top) and system 4 (bottom).}
\label{Fig6}
\end{figure}

Applying this analysis to the configurations generated during the umbrella sampling windows, leads to the determination of the evolution of the size of the droplet during the nucleation process. Furthermore, by keeping track of the identity (either $Ar$ or $Kr$) of the atoms belonging to the droplet, we can also shed light on the chemical selectivity during the formation of the cluster. This means that the interplay between droplet size and composition during the nucleation process can be directly accessed during the simulations, and, in turn, shed light on the departure in composition of the critical droplet from the bulk composition~\cite{zeng1991binary,braun2014molecular} and the possible onset of phase separation in partially miscible mixtures~\cite{talanquer1995nucleation,ten1998numerical,napari1999gas}.

We show on the left of Fig.~\ref{Fig6}, the results obtained for the two supersaturations of systems 1 and 2. In both cases, throughout the nucleation process, the overall size of the droplet is shown to gradually increase as the entropy of the system decreases. The critical size of the droplet is found to be larger at low supersaturation ($N_c=713\pm60$ for system 1 and $N_c=323\pm30$ for system 2). It is also reached for a lower value of the entropy at low supersaturation ($S_c=0.103$~kJ/mol/K for system 1) than at high supersaturation ($S_c=0.1055$~kJ/mol/K for system 2), in line with the results obtained for the free energy barrier in Fig.~\ref{Fig3}. A closer inspection of the variation for the number of each type of atoms show that the composition of the droplet does not remain the same during the entire nucleation process. For small sizes (high entropy), the droplet is richer in $Ar$ with $55\pm5$~\% of Ar atoms in the droplet at $S=0.1073$~kJ/mol/K for system 1. There is then a crossover at $S=0.1067$~kJ/mol/K for which the two types of atoms are equally present. As the entropy further decreases, the droplet becomes richer in $Kr$ atoms (with a fraction of $48\pm4$~\% in $Ar$ for a droplet of a critical size). Looking at the composition of the droplet past the critical size, we find that the fraction of $Ar$ atoms in the droplet increases again and reaches $49\pm4$~\% for $S=0.102$~kJ/mol/K. The same mechanism is observed for system 2, with a crossover point located at $S=0.1068$~kJ/mol/K and a fraction of $Ar$ atoms of $47\pm4$~\% in the critical droplet. These fluctuations in the nucleus composition through the nucleation process, as well as the departure of the composition of the droplet from the composition of the liquid phase, are consistent with prior work on nucleation, based on a revision of the classical nucleation theory~\cite{wilemski1987revised}, on classical density functional theory calculations~\cite{zeng1991binary}, on molecular simulations~\cite{ten1998numerical,yoo2001monte,braun2014molecular}, or on approaches based on macroscopic kinetics~\cite{alekseechkin2015thermodynamics}. Our results indicate that the mole fraction in $Ar$ in the nucleus to be less than for the bulk by $0.08$ for System 1 and of $0.03$ for System 2 with respect to the bulk composition. These deviations, which are moderate due the almost ideal nature of the $Ar-Kr$ mixture as noted by Zeng and Oxtoby~\cite{zeng1991binary}, did not give rise to a significant phase separation effect as reported in simulations of more strongly asymmetric mixtures~\cite{ten1998numerical,napari1999density}.

Looking now at the evolution of the size of the droplet as a function of entropy for the other set of conditions ($x_{Ar}=0.75$), we see that the size of the droplet steadily increases as the entropy of the system decreases. The droplet reaches a size of $629\pm90$ atoms for the critical droplet in the case of system 3, and a size of $273\pm50$ for system 4. As for the previous system, the critical droplet has formed once the entropy has reached a critical value $S_c=0.100$~kJ/mol/K for system 3, a value that is notably lower than its counterpart of $0.103$ for system 4. This reflects the fact that the range for the entropies spanned during the nucleation event becomes narrower and narrower as the supersaturation is increased. Unlike for systems 1 and 2, we do not observe any crossover between the mole fractions in $Ar$ and $Kr$ in the droplet, given the very large fraction of $Ar$ in systems 3 and 4. As for systems 1 and 2, we observe, however, fluctuations in the composition of the droplet with the fraction of $Ar$ atoms in the droplet varying between $60$~\% and $80$~\% during nucleation. 

\subsection{$CO_2-C_2H_6$ mixture}

\begin{figure}
\begin{center}
\includegraphics*[width=8cm]{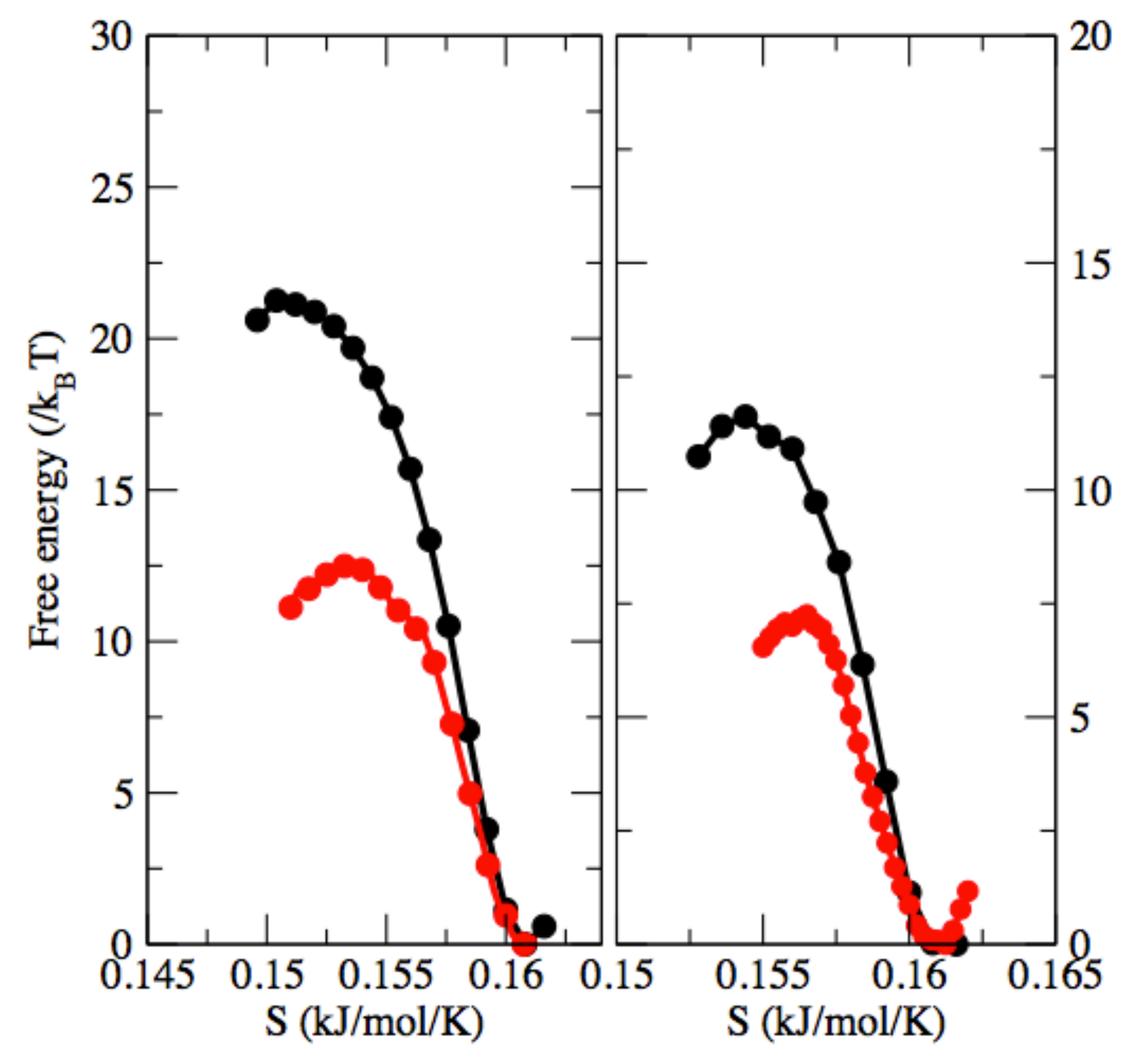}
\end{center}
\caption{Free energy barriers of nucleation for the $CO_{2}-C_2H_6$ mixture at $T=263.15$~K. (Left panel) System 5 (black) and system 6 (red) for $x_{CO_2}=0.053$. (Right panel) System 7 (black) and system 8 (red) for $x_{CO_2}=0.097$}
\label{Fig7}
\end{figure}

Turning to the results obtained for the $CO_2-C_2H_6$ mixture, we plot in Fig.~\ref{Fig7}, the free energy barrier of nucleation obtained for systems 5 and 6. For a liquid fraction of $x_{CO_2}=0.053$ (left panel of Fig.~\ref{Fig7}), we observe that increasing the supersaturation leads to a decrease in the height of the free energy barrier of nucleation from $21\pm2~k_BT$ (system 5) to $12\pm1~k_BT$ (system 6). Similarly, for a mole fraction of $x_{CO_2}=0.097$, a lower supersaturation results in a barrier of $12\pm1~k_BT$ (system 7), while a higher supersaturation yields a free energy of nucleation of $7\pm1~k_BT$ (system 8). The range of entropies spanned along the nucleation pathway is also found to depend strongly on supersaturation, and becomes broader at lower supersaturations. For instance, for system 5, we obtain an entropy for the parent (supersaturated vapor) phase of $0.161$~kJ/mol/K and an entropy of the system of $S_c=0.151$~kJ/mol/K, when a droplet of a critical size has formed. Increasing the supersaturation (system 6) leads to a narrower entropy range, most notably as a result of a decrease in the entropy $S_c=0.154$~kJ/mol/K for which the critical size of the droplet is reached. Similar conclusions apply for systems 7 and 8. We find a narrowing of the entropy range spanned during nucleation at high supersaturation, with the top of the free energy barrier droplet reached for a higher entropy ($S_c=0.157$~kJ/mol/K) than at low supersaturation ($S_c=0.155$~kJ/mol/K). The higher entropy required to form a droplet of a critical size at high supersaturation, together with the lower free energy barrier of nucleation obtained at high supersaturation, are most likely due to the smaller size of the critical droplet at high supersaturation, a point that we discuss in the next section in our analysis of the relation between size, selectivity and entropy during the nucleation process.

\begin{figure}
\begin{center}
\includegraphics*[width=5cm]{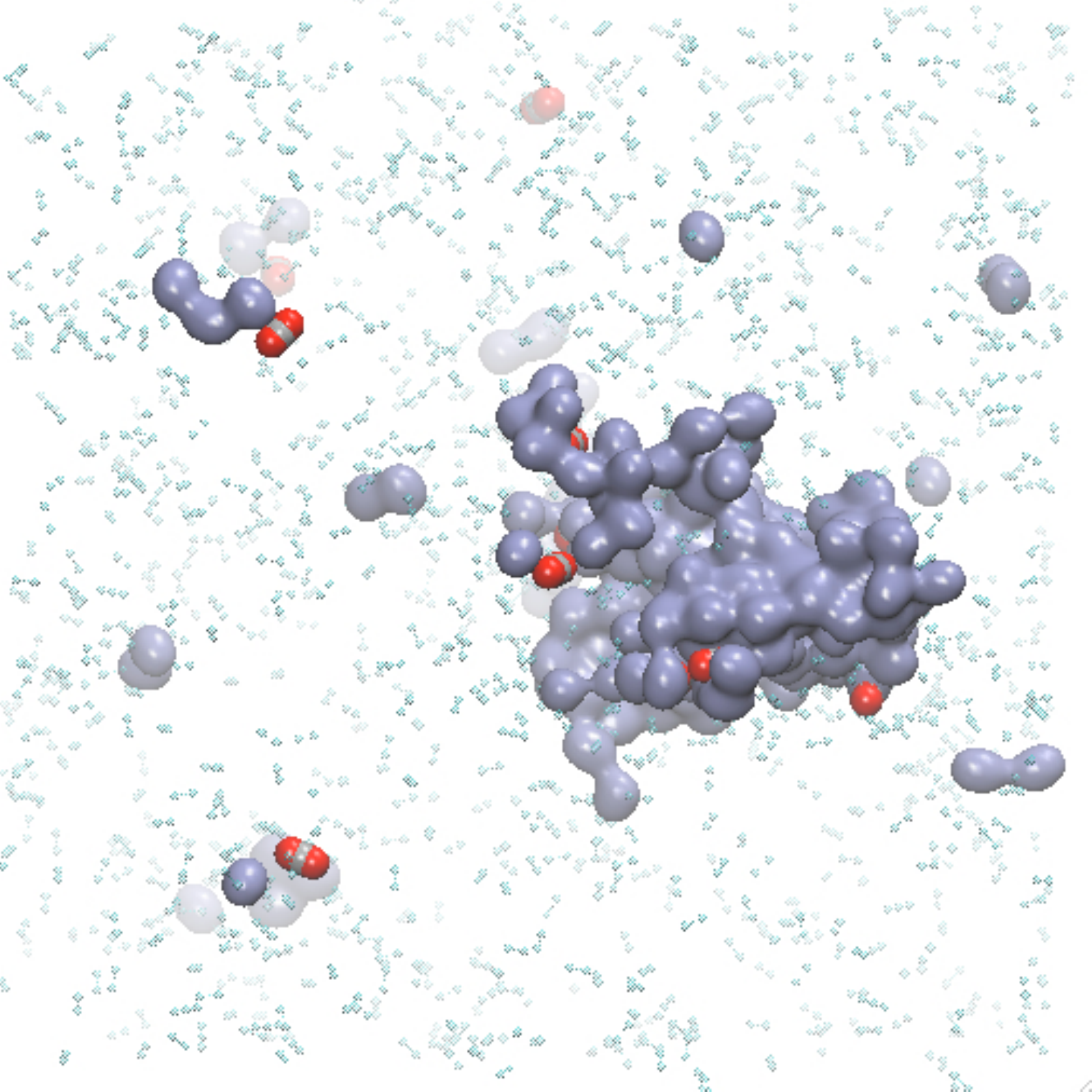}
\includegraphics*[width=5cm]{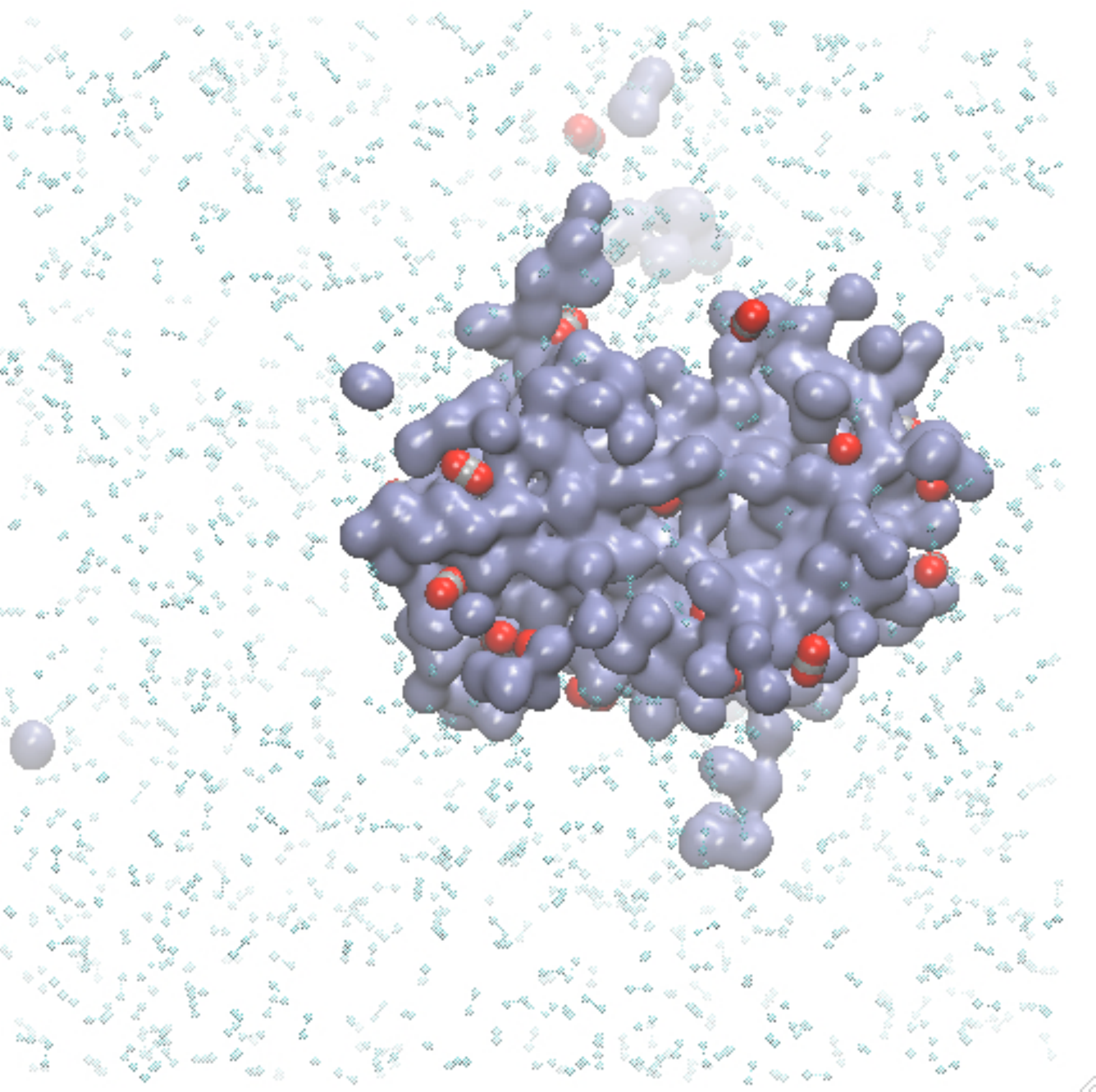}
\includegraphics*[width=5cm]{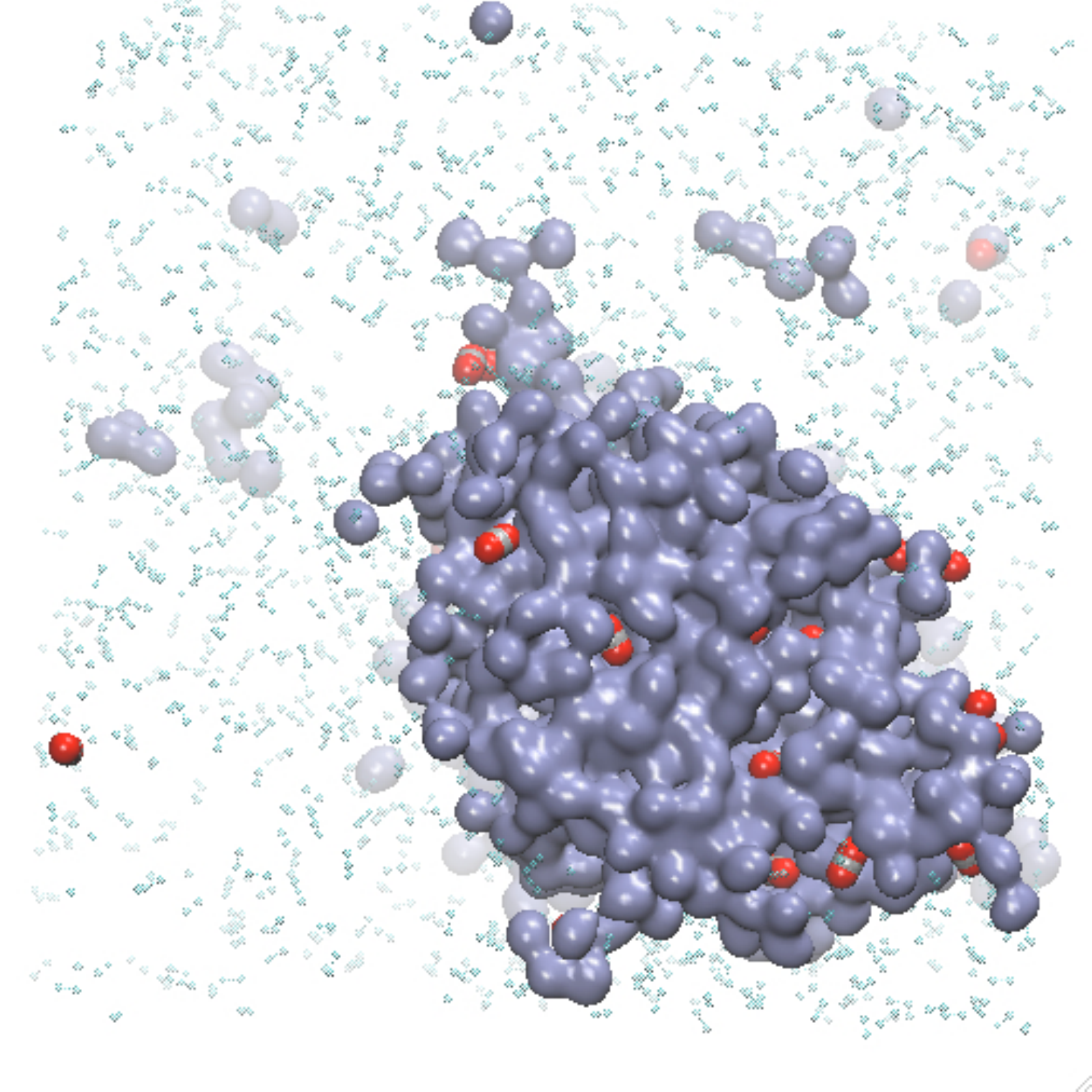}
\end{center}
\caption{$C_2H_6-CO_2$ mixture: Snapshots of system 5 during the nucleation process for a target entropy of $0.157$~kJ/mol/K (left), $0.154$~kJ/mol/K (middle) and $0.151$~kJ/mol/K (right). Molecules with a liquid-like environment are shown with larger spheres, with the $CO_2$ molecules identified through their $O$ atoms shown in red.}
\label{Fig8}
\end{figure}

We show in Fig.~\ref{Fig8} snapshots of configurations of system 5, obtained throughout the nucleation process. As for the $Ar-Kr$ system, we start from a metastable supersaturated vapor and carry out umbrella sampling windows with decreasing values for the target entropy. During the first few umbrella sampling windows, the target entropy is high enough, so that the droplet is fairly small (see on the left of Fig.~\ref{Fig8} for $S_{0,i}=0.157$~kJ/kg/K). Then, as $S_{0,i}$ is decreased further, the droplet starts to become larger and larger, and eventually reaches its critical size for $S_{0,i}=0.151$~kJ/kg/K. This shows that decreasing the value for the target entropy not only increases the density of the system, but also results in an increased level of organization with the formation of the liquid droplet.

\begin{figure}
\begin{center}
\includegraphics*[width=8cm]{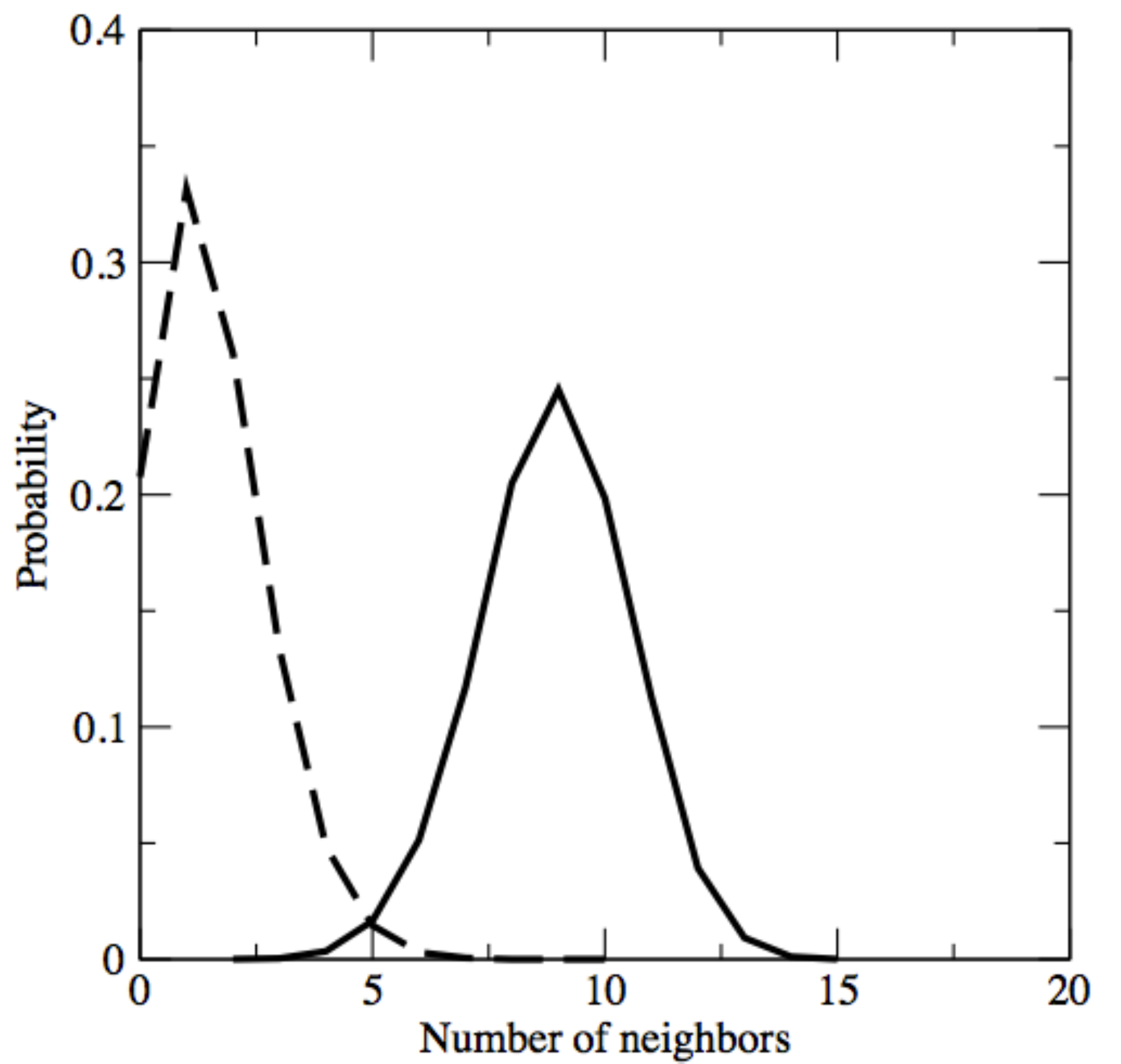}
\end{center}
\caption{$CO_2-C_2H_6$ mixture: Distributions for the number of neighbors within a distance of $6.4$~\AA~for the liquid (solid line) and for the vapor (dashed line) at coexistence for $T=263.15$~K.}
\label{Fig9}
\end{figure}

As for the Ar-Kr mixture, Fig.~\ref{Fig7} shows a projection of the multi-dimensional free energy surface in the (entropy, free energy) plane. We pursue our analysis by characterizing the interdependence between the entropy S, the droplet size and its composition. We start by determining which molecules in the system have a liquid-like environment and, as such, belong to the droplet. We determine, for each molecule, the distributions, shown in Fig.~\ref{Fig9}, for the number of neighbors for the vapor and the liquid. Neighboring molecules are defined as being separated by a distance (between the centers of mass of the 2 neighboring molecules) less than $6.4$~\AA. The distributions obtained for the vapor and liquid reach their maxima for very different numbers of neighbors (for, on average, a single neighbor in the case of the vapor and for $9$ neighbors in the case of the liquid). This allows us to define a molecule as having a liquid-like environment. and thus belonging to the droplet, if it has $6$ or more neighbors within a spherical shell of $6.4$~\AA.

\begin{figure}
\begin{center}
\includegraphics*[width=10cm]{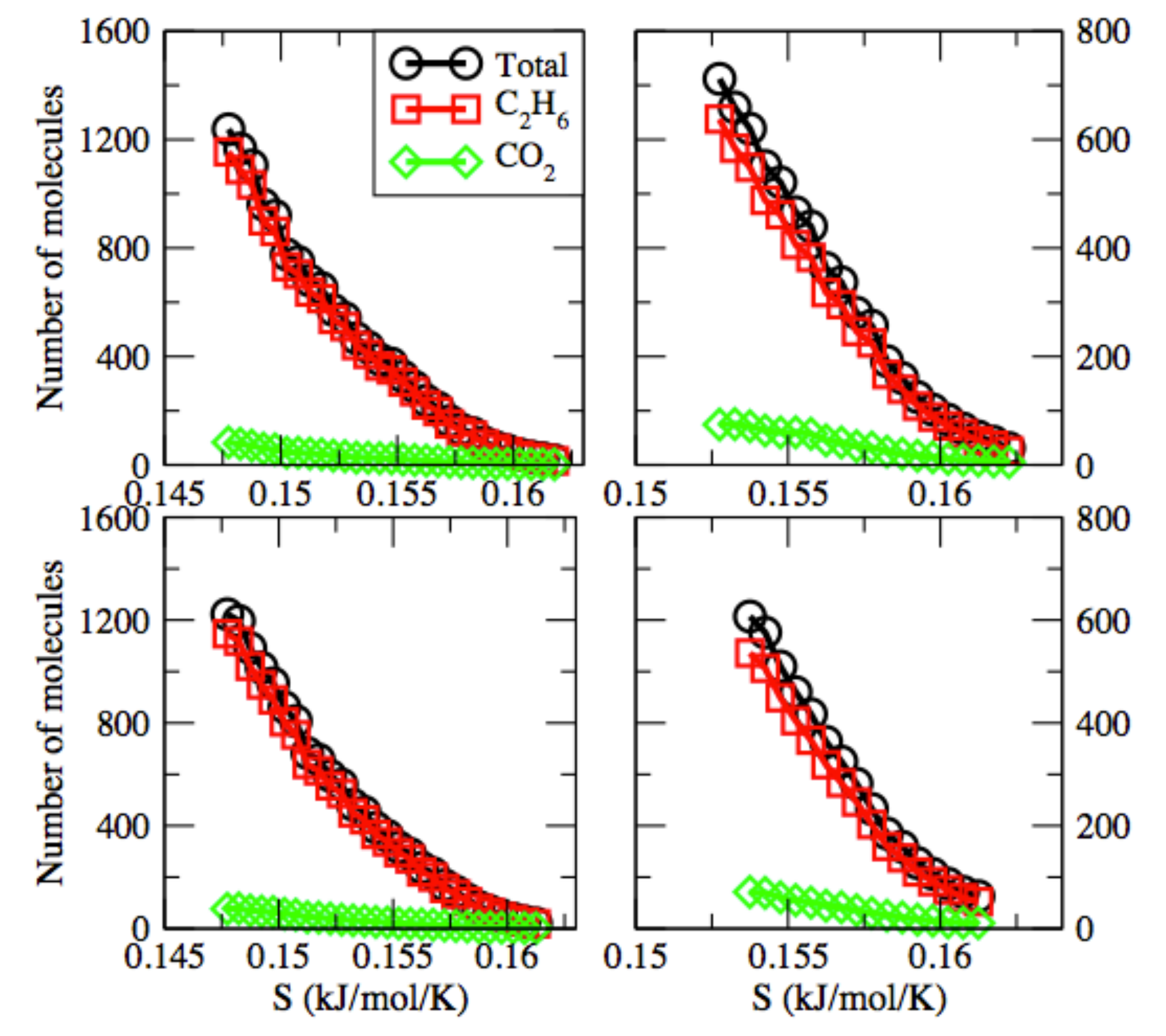}
\end{center}
\caption{$C_2H_6-CO_2$ mixture. (Left panel) Variation of the total number of molecules in the cluster and of the number of molecules for $C_2H_6$ and $CO_2$ at $x_{CO_2}=0.053$ for system 5 (top) and system 6 (bottom). (Right panel) Total number of molecules in the cluster, number of molecules for $C_2H_6$ and $CO_2$ against $S$ at $x_{CO_2}=0.097$ for system 7 (top) and system 8 (bottom).}
\label{Fig10}
\end{figure}

We now move on to the analysis of the size of the droplet as a function of the entropy of the system throughout the nucleation process. Fig.~\ref{Fig10} shows that, for all systems, the total number of molecules within the cluster increases smoothly as the entropy of the system decreases. Furthermore, we find that the size of the critical droplet decreases as supersaturation is increased. For a liquid mole fraction of $x_{CO_2}=0.053$, the critical size for the droplet is of $N_c=715\pm35$ molecules for system 5 and of $N_c=422\pm30$ molecules for system 6. The smaller size of the critical droplet at high supersaturation (system 6) accounts for the higher value of the entropy for which the system reaches the top of the free energy barrier. Similarly, when the liquid mole fraction $x_{CO_2}=0.097$, the critical size is of $N_c=494\pm28$ molecules for a low supersaturation (system 7) and of $N_c=303\pm21$ molecules for a high supersaturation (system 8). The smaller critical size and higher critical entropy obtained at the higher supersaturation are, once again, found to be consistent with the free energy plot of Fig.~\ref{Fig7}. Turning to the composition of the droplet,  we find that $C_2H_6$ remains predominant throughout the nucleation process for all systems (see Fig.~\ref{Fig10}). We also find, however, that the composition of the nucleus depends on its size, as nucleation starts with the formation of a droplet that has a higher $CO_2$ mole fraction than the bulk. It is around $8$~\% for systems 5 and 6 for droplets containing a total of $50-100$ molecules. Similarly, considering the same droplet sizes, it is of about $17$~\% for systems 7 and 8. The fraction of $CO_2$ then decreases as the size of the droplet increases. For droplets of a critical size, the fraction of $CO_2$ is of $6$~\% for systems 5 and 6, while it is of $13$~\% for systems 7 and 8. Despite the small sizes of the critical droplets, which contain only a few hundred of molecules, the fractions in the critical droplets are reasonably close to the $CO_2$ mole fraction of the liquid, and the departure from the bulk compositions ($5.3$~\% of $CO_2$ for systems 5 and 6, and $9.7$~\% of $CO_2$ for systems 7 and 8) is small. As for binary mixtures of atoms, the departures in the droplet composition with respect to that of the bulk are consistent with prior simulations of droplet nucleation in binary molecular systems (see e.g. recent simulations of the methane-nonane system~\cite{braun2014molecular}).

\section{Conclusion}
In this work, we propose a new simulation method to study the nucleation process in binary mixtures of atomic fluids ($Ar-Kr$) and of molecular fluids ($C_2H_6-CO_2$). The method is based on driving the formation of a liquid droplet through a series of umbrella sampling simulations where the bias potential is a function of the entropy $S$ of the system. The resulting approach is implemented within the grand-canonical ensemble and, since the entropy serves as the reaction coordinate for the nucleation process, is called $\mu_1 \mu_2 VT-S$. The application of the method to the formation of liquid droplets in binary mixtures sheds light on the interplay between the size of the droplet, its composition and the supersaturation at which the nucleation process occurs. Our findings show that, at low supersaturation, the range of entropies spanned by the nucleation process becomes broader, as a result of the combined effect of the larger entropy of the metastable supersaturated vapor (parent phase) and of the lower entropy associated with the configurations of the system that contain a liquid droplet of a critical size. These simulations allow us to characterize the critical droplet in terms of a critical value reached by the entropy at the top of the free energy barrier of nucleation. We are also able to obtain the free energy profile along the entropic pathway underlying the formation of the liquid droplet and to calculate the free energy of nucleation as a function of the supersaturation and chemical composition of the system. The analysis of the composition of the droplet shows that the mole fractions fluctuate throughout the nucleation process and depart from the composition of the bulk. This departure is however found to become less and less significant as the size of the droplet increases and its composition starts to conform more and more to that predicted by thermodynamics. Finally, while the $\mu_1\mu_2 VT-S$ does not yield directly the nucleation rate, the method allows to generate and stabilize configurations of the system close to the top of the free energy barrier. However, as discussed in previous work~\cite{ten1999numerical}, the nucleation rate can be obtained by carrying out additional molecular molecular dynamics simulations, using configurations close to the top of the free energy barrier as a starting point, and following the Bennett-Chandler scheme~\cite{bennett1977algorithms,chandler1978statistical,carter1989constrained} to determine the kinetics of the process. Alternatively, the thereshold method of Yasuoka and Matsumoto can also be used to determine the nucleation rate~\cite{yasuoka1998molecular}.

{\bf Acknowledgements}
Partial funding for this research was provided by NSF through CAREER award DMR-1052808.\\

\bibliography{muVTS_II}

\begin{thebibliography}{105}
\expandafter\ifx\csname natexlab\endcsname\relax\def\natexlab#1{#1}\fi
\expandafter\ifx\csname bibnamefont\endcsname\relax
  \def\bibnamefont#1{#1}\fi
\expandafter\ifx\csname bibfnamefont\endcsname\relax
  \def\bibfnamefont#1{#1}\fi
\expandafter\ifx\csname citenamefont\endcsname\relax
  \def\citenamefont#1{#1}\fi
\expandafter\ifx\csname url\endcsname\relax
  \def\url#1{\texttt{#1}}\fi
\expandafter\ifx\csname urlprefix\endcsname\relax\def\urlprefix{URL }\fi
\providecommand{\bibinfo}[2]{#2}
\providecommand{\eprint}[2][]{\url{#2}}

\bibitem[{\citenamefont{Yasuoka and Matsumoto}(1998)}]{yasuoka1998molecular}
\bibinfo{author}{\bibfnamefont{K.}~\bibnamefont{Yasuoka}} \bibnamefont{and}
  \bibinfo{author}{\bibfnamefont{M.}~\bibnamefont{Matsumoto}},
  \bibinfo{journal}{J. Chem. Phys.} \textbf{\bibinfo{volume}{109}},
  \bibinfo{pages}{8451} (\bibinfo{year}{1998}).

\bibitem[{\citenamefont{Oxtoby}(1992)}]{oxtoby1992homogeneous}
\bibinfo{author}{\bibfnamefont{D.~W.} \bibnamefont{Oxtoby}},
  \bibinfo{journal}{J. Phys. Condens. Matter} \textbf{\bibinfo{volume}{4}},
  \bibinfo{pages}{7627} (\bibinfo{year}{1992}).

\bibitem[{\citenamefont{Shen and Debenedetti}(1999)}]{shen1999computational}
\bibinfo{author}{\bibfnamefont{V.~K.} \bibnamefont{Shen}} \bibnamefont{and}
  \bibinfo{author}{\bibfnamefont{P.~G.} \bibnamefont{Debenedetti}},
  \bibinfo{journal}{J. Chem. Phys.} \textbf{\bibinfo{volume}{111}},
  \bibinfo{pages}{3581} (\bibinfo{year}{1999}).

\bibitem[{\citenamefont{Weakliem and Reiss}(1993)}]{weakliem1993toward}
\bibinfo{author}{\bibfnamefont{C.~L.} \bibnamefont{Weakliem}} \bibnamefont{and}
  \bibinfo{author}{\bibfnamefont{H.}~\bibnamefont{Reiss}}, \bibinfo{journal}{J.
  Chem. Phys.} \textbf{\bibinfo{volume}{99}}, \bibinfo{pages}{5374}
  (\bibinfo{year}{1993}).

\bibitem[{\citenamefont{Schenter et~al.}(1999)\citenamefont{Schenter, Kathmann,
  and Garrett}}]{schenter1999dynamical}
\bibinfo{author}{\bibfnamefont{G.~K.} \bibnamefont{Schenter}},
  \bibinfo{author}{\bibfnamefont{S.~M.} \bibnamefont{Kathmann}},
  \bibnamefont{and} \bibinfo{author}{\bibfnamefont{B.~C.}
  \bibnamefont{Garrett}}, \bibinfo{journal}{Phys. Rev. Lett.}
  \textbf{\bibinfo{volume}{82}}, \bibinfo{pages}{3484} (\bibinfo{year}{1999}).

\bibitem[{\citenamefont{Zeng and Oxtoby}(1991{\natexlab{a}})}]{zeng1991gas}
\bibinfo{author}{\bibfnamefont{X.~C.} \bibnamefont{Zeng}} \bibnamefont{and}
  \bibinfo{author}{\bibfnamefont{D.~W.} \bibnamefont{Oxtoby}},
  \bibinfo{journal}{J. Chem. Phys.} \textbf{\bibinfo{volume}{94}},
  \bibinfo{pages}{4472} (\bibinfo{year}{1991}{\natexlab{a}}).

\bibitem[{\citenamefont{Yi et~al.}(2002)\citenamefont{Yi, Poulikakos, Walther,
  and Yadigaroglu}}]{yi2002molecular}
\bibinfo{author}{\bibfnamefont{P.}~\bibnamefont{Yi}},
  \bibinfo{author}{\bibfnamefont{D.}~\bibnamefont{Poulikakos}},
  \bibinfo{author}{\bibfnamefont{J.}~\bibnamefont{Walther}}, \bibnamefont{and}
  \bibinfo{author}{\bibfnamefont{G.}~\bibnamefont{Yadigaroglu}},
  \bibinfo{journal}{Int. J. Heat Mass Tran.} \textbf{\bibinfo{volume}{45}},
  \bibinfo{pages}{2087} (\bibinfo{year}{2002}).

\bibitem[{\citenamefont{Kinjo et~al.}(1999)\citenamefont{Kinjo, Ohguchi,
  Yasuoka, and Matsumoto}}]{kinjo1999computer}
\bibinfo{author}{\bibfnamefont{T.}~\bibnamefont{Kinjo}},
  \bibinfo{author}{\bibfnamefont{K.}~\bibnamefont{Ohguchi}},
  \bibinfo{author}{\bibfnamefont{K.}~\bibnamefont{Yasuoka}}, \bibnamefont{and}
  \bibinfo{author}{\bibfnamefont{M.}~\bibnamefont{Matsumoto}},
  \bibinfo{journal}{Comput. Mater. Sci.} \textbf{\bibinfo{volume}{14}},
  \bibinfo{pages}{138} (\bibinfo{year}{1999}).

\bibitem[{\citenamefont{Toxvaerd}(2001)}]{toxvaerd2001molecular}
\bibinfo{author}{\bibfnamefont{S.}~\bibnamefont{Toxvaerd}},
  \bibinfo{journal}{J. Chem. Phys.} \textbf{\bibinfo{volume}{115}},
  \bibinfo{pages}{8913} (\bibinfo{year}{2001}).

\bibitem[{\citenamefont{Ford}(1996)}]{ford1996thermodynamic}
\bibinfo{author}{\bibfnamefont{I.}~\bibnamefont{Ford}}, \bibinfo{journal}{J.
  Chem. Phys.} \textbf{\bibinfo{volume}{105}}, \bibinfo{pages}{8324}
  (\bibinfo{year}{1996}).

\bibitem[{\citenamefont{Talanquer and
  Oxtoby}(1995{\natexlab{a}})}]{talanquer1995density}
\bibinfo{author}{\bibfnamefont{V.}~\bibnamefont{Talanquer}} \bibnamefont{and}
  \bibinfo{author}{\bibfnamefont{D.}~\bibnamefont{Oxtoby}},
  \bibinfo{journal}{J. Phys. Chem.} \textbf{\bibinfo{volume}{99}},
  \bibinfo{pages}{2865} (\bibinfo{year}{1995}{\natexlab{a}}).

\bibitem[{\citenamefont{Reiss et~al.}(1990)\citenamefont{Reiss, Tabazadeh, and
  Talbot}}]{reiss1990molecular}
\bibinfo{author}{\bibfnamefont{H.}~\bibnamefont{Reiss}},
  \bibinfo{author}{\bibfnamefont{A.}~\bibnamefont{Tabazadeh}},
  \bibnamefont{and} \bibinfo{author}{\bibfnamefont{J.}~\bibnamefont{Talbot}},
  \bibinfo{journal}{J. Chem. Phys.} \textbf{\bibinfo{volume}{92}},
  \bibinfo{pages}{1266} (\bibinfo{year}{1990}).

\bibitem[{\citenamefont{Kalikmanov and
  Van~Dongen}(1995)}]{kalikmanov1995semiphenomenological}
\bibinfo{author}{\bibfnamefont{V.}~\bibnamefont{Kalikmanov}} \bibnamefont{and}
  \bibinfo{author}{\bibfnamefont{M.}~\bibnamefont{Van~Dongen}},
  \bibinfo{journal}{J. Chem. Phys.} \textbf{\bibinfo{volume}{103}},
  \bibinfo{pages}{4250} (\bibinfo{year}{1995}).

\bibitem[{\citenamefont{Horsch et~al.}(2008)\citenamefont{Horsch, Vrabec, and
  Hasse}}]{horsch2008modification}
\bibinfo{author}{\bibfnamefont{M.}~\bibnamefont{Horsch}},
  \bibinfo{author}{\bibfnamefont{J.}~\bibnamefont{Vrabec}}, \bibnamefont{and}
  \bibinfo{author}{\bibfnamefont{H.}~\bibnamefont{Hasse}},
  \bibinfo{journal}{Phys. Rev. E} \textbf{\bibinfo{volume}{78}},
  \bibinfo{pages}{011603} (\bibinfo{year}{2008}).

\bibitem[{\citenamefont{Neimark and Vishnyakov}(2005)}]{neimark2005birth}
\bibinfo{author}{\bibfnamefont{A.~V.} \bibnamefont{Neimark}} \bibnamefont{and}
  \bibinfo{author}{\bibfnamefont{A.}~\bibnamefont{Vishnyakov}},
  \bibinfo{journal}{J. Chem. Phys.} \textbf{\bibinfo{volume}{122}},
  \bibinfo{pages}{054707} (\bibinfo{year}{2005}).

\bibitem[{\citenamefont{Oxtoby and Evans}(1988)}]{oxtoby1988nonclassical}
\bibinfo{author}{\bibfnamefont{D.~W.} \bibnamefont{Oxtoby}} \bibnamefont{and}
  \bibinfo{author}{\bibfnamefont{R.}~\bibnamefont{Evans}}, \bibinfo{journal}{J.
  Chem. Phys.} \textbf{\bibinfo{volume}{89}}, \bibinfo{pages}{7521}
  (\bibinfo{year}{1988}).

\bibitem[{\citenamefont{Lutsko}(2008)}]{lutsko2008density}
\bibinfo{author}{\bibfnamefont{J.~F.} \bibnamefont{Lutsko}},
  \bibinfo{journal}{J. Chem. Phys.} \textbf{\bibinfo{volume}{129}},
  \bibinfo{pages}{244501} (\bibinfo{year}{2008}).

\bibitem[{\citenamefont{Wang et~al.}(2008)\citenamefont{Wang, Valeriani, and
  Frenkel}}]{wang2008homogeneous}
\bibinfo{author}{\bibfnamefont{Z.-J.} \bibnamefont{Wang}},
  \bibinfo{author}{\bibfnamefont{C.}~\bibnamefont{Valeriani}},
  \bibnamefont{and} \bibinfo{author}{\bibfnamefont{D.}~\bibnamefont{Frenkel}},
  \bibinfo{journal}{J. Phys. Chem. B} \textbf{\bibinfo{volume}{113}},
  \bibinfo{pages}{3776} (\bibinfo{year}{2008}).

\bibitem[{\citenamefont{Ten~Wolde et~al.}(1999)\citenamefont{Ten~Wolde,
  Ruiz-Montero, and Frenkel}}]{ten1999numerical}
\bibinfo{author}{\bibfnamefont{P.~R.} \bibnamefont{Ten~Wolde}},
  \bibinfo{author}{\bibfnamefont{M.~J.} \bibnamefont{Ruiz-Montero}},
  \bibnamefont{and} \bibinfo{author}{\bibfnamefont{D.}~\bibnamefont{Frenkel}},
  \bibinfo{journal}{J. Chem. Phys.} \textbf{\bibinfo{volume}{110}},
  \bibinfo{pages}{1591} (\bibinfo{year}{1999}).

\bibitem[{\citenamefont{Gonzalez et~al.}(2015)\citenamefont{Gonzalez, Abascal,
  Valeriani, and Bresme}}]{gonzalez2015bubble}
\bibinfo{author}{\bibfnamefont{M.~A.} \bibnamefont{Gonzalez}},
  \bibinfo{author}{\bibfnamefont{J.~L.} \bibnamefont{Abascal}},
  \bibinfo{author}{\bibfnamefont{C.}~\bibnamefont{Valeriani}},
  \bibnamefont{and} \bibinfo{author}{\bibfnamefont{F.}~\bibnamefont{Bresme}},
  \bibinfo{journal}{J. Chem. Phys.} \textbf{\bibinfo{volume}{142}},
  \bibinfo{pages}{154903} (\bibinfo{year}{2015}).

\bibitem[{\citenamefont{Loeffler et~al.}(2015)\citenamefont{Loeffler, Sepehri,
  and Chen}}]{loeffler2015improved}
\bibinfo{author}{\bibfnamefont{T.~D.} \bibnamefont{Loeffler}},
  \bibinfo{author}{\bibfnamefont{A.}~\bibnamefont{Sepehri}}, \bibnamefont{and}
  \bibinfo{author}{\bibfnamefont{B.}~\bibnamefont{Chen}}, \bibinfo{journal}{J.
  Chem. Theory Comput.} \textbf{\bibinfo{volume}{11}}, \bibinfo{pages}{4023}
  (\bibinfo{year}{2015}).

\bibitem[{\citenamefont{Sosso et~al.}(2016)\citenamefont{Sosso, Chen, Cox,
  Fitzner, Pedevilla, Zen, and Michaelides}}]{sosso2016crystal}
\bibinfo{author}{\bibfnamefont{G.~C.} \bibnamefont{Sosso}},
  \bibinfo{author}{\bibfnamefont{J.}~\bibnamefont{Chen}},
  \bibinfo{author}{\bibfnamefont{S.~J.} \bibnamefont{Cox}},
  \bibinfo{author}{\bibfnamefont{M.}~\bibnamefont{Fitzner}},
  \bibinfo{author}{\bibfnamefont{P.}~\bibnamefont{Pedevilla}},
  \bibinfo{author}{\bibfnamefont{A.}~\bibnamefont{Zen}}, \bibnamefont{and}
  \bibinfo{author}{\bibfnamefont{A.}~\bibnamefont{Michaelides}},
  \bibinfo{journal}{Chem. Rev.}  (\bibinfo{year}{2016}).

\bibitem[{\citenamefont{Xu et~al.}(2015)\citenamefont{Xu, Lan, Peng, Wen, and
  Ma}}]{xu2015effect}
\bibinfo{author}{\bibfnamefont{W.}~\bibnamefont{Xu}},
  \bibinfo{author}{\bibfnamefont{Z.}~\bibnamefont{Lan}},
  \bibinfo{author}{\bibfnamefont{B.}~\bibnamefont{Peng}},
  \bibinfo{author}{\bibfnamefont{R.}~\bibnamefont{Wen}}, \bibnamefont{and}
  \bibinfo{author}{\bibfnamefont{X.}~\bibnamefont{Ma}}, \bibinfo{journal}{J.
  Chem. Phys.} \textbf{\bibinfo{volume}{142}}, \bibinfo{pages}{054701}
  (\bibinfo{year}{2015}).

\bibitem[{\citenamefont{Keasler and Siepmann}(2015)}]{keasler2015understanding}
\bibinfo{author}{\bibfnamefont{S.~J.} \bibnamefont{Keasler}} \bibnamefont{and}
  \bibinfo{author}{\bibfnamefont{J.~I.} \bibnamefont{Siepmann}},
  \bibinfo{journal}{J. Chem. Phys.} \textbf{\bibinfo{volume}{143}},
  \bibinfo{pages}{164516} (\bibinfo{year}{2015}).

\bibitem[{\citenamefont{Wilhelmsen et~al.}(2015)\citenamefont{Wilhelmsen,
  Trinh, Kjelstrup, and Bedeaux}}]{wilhelmsen2015influence}
\bibinfo{author}{\bibfnamefont{{\O}.}~\bibnamefont{Wilhelmsen}},
  \bibinfo{author}{\bibfnamefont{T.~T.} \bibnamefont{Trinh}},
  \bibinfo{author}{\bibfnamefont{S.}~\bibnamefont{Kjelstrup}},
  \bibnamefont{and} \bibinfo{author}{\bibfnamefont{D.}~\bibnamefont{Bedeaux}},
  \bibinfo{journal}{J. Phys. Chem. C} \textbf{\bibinfo{volume}{119}},
  \bibinfo{pages}{8160} (\bibinfo{year}{2015}).

\bibitem[{\citenamefont{van Meel et~al.}(2015)\citenamefont{van Meel, Liu, and
  Frenkel}}]{van2015mechanism}
\bibinfo{author}{\bibfnamefont{J.}~\bibnamefont{van Meel}},
  \bibinfo{author}{\bibfnamefont{Y.}~\bibnamefont{Liu}}, \bibnamefont{and}
  \bibinfo{author}{\bibfnamefont{D.}~\bibnamefont{Frenkel}},
  \bibinfo{journal}{Mol. Phys.} \textbf{\bibinfo{volume}{113}},
  \bibinfo{pages}{2742} (\bibinfo{year}{2015}).

\bibitem[{\citenamefont{Hale}(1986)}]{hale1986application}
\bibinfo{author}{\bibfnamefont{B.~N.} \bibnamefont{Hale}},
  \bibinfo{journal}{Phys. Rev. A} \textbf{\bibinfo{volume}{33}},
  \bibinfo{pages}{4156} (\bibinfo{year}{1986}).

\bibitem[{\citenamefont{Hale}(2005)}]{hale2005temperature}
\bibinfo{author}{\bibfnamefont{B.~N.} \bibnamefont{Hale}}, \bibinfo{journal}{J.
  Chem. Phys.} \textbf{\bibinfo{volume}{122}}, \bibinfo{pages}{204509}
  (\bibinfo{year}{2005}).

\bibitem[{\citenamefont{Hale and Thomason}(2010)}]{hale2010scaled}
\bibinfo{author}{\bibfnamefont{B.~N.} \bibnamefont{Hale}} \bibnamefont{and}
  \bibinfo{author}{\bibfnamefont{M.}~\bibnamefont{Thomason}},
  \bibinfo{journal}{Phys. Rev. Lett.} \textbf{\bibinfo{volume}{105}},
  \bibinfo{pages}{046101} (\bibinfo{year}{2010}).

\bibitem[{\citenamefont{Yuhara et~al.}(2015)\citenamefont{Yuhara, Barnes, Suh,
  Knott, Beckham, Yasuoka, Wu, and Sum}}]{yuhara2015nucleation}
\bibinfo{author}{\bibfnamefont{D.}~\bibnamefont{Yuhara}},
  \bibinfo{author}{\bibfnamefont{B.~C.} \bibnamefont{Barnes}},
  \bibinfo{author}{\bibfnamefont{D.}~\bibnamefont{Suh}},
  \bibinfo{author}{\bibfnamefont{B.~C.} \bibnamefont{Knott}},
  \bibinfo{author}{\bibfnamefont{G.~T.} \bibnamefont{Beckham}},
  \bibinfo{author}{\bibfnamefont{K.}~\bibnamefont{Yasuoka}},
  \bibinfo{author}{\bibfnamefont{D.~T.} \bibnamefont{Wu}}, \bibnamefont{and}
  \bibinfo{author}{\bibfnamefont{A.~K.} \bibnamefont{Sum}},
  \bibinfo{journal}{Faraday Discuss.} \textbf{\bibinfo{volume}{179}},
  \bibinfo{pages}{463} (\bibinfo{year}{2015}).

\bibitem[{\citenamefont{Lauricella et~al.}(2015)\citenamefont{Lauricella,
  Meloni, Liang, English, Kusalik, and Ciccotti}}]{lauricella2015clathrate}
\bibinfo{author}{\bibfnamefont{M.}~\bibnamefont{Lauricella}},
  \bibinfo{author}{\bibfnamefont{S.}~\bibnamefont{Meloni}},
  \bibinfo{author}{\bibfnamefont{S.}~\bibnamefont{Liang}},
  \bibinfo{author}{\bibfnamefont{N.~J.} \bibnamefont{English}},
  \bibinfo{author}{\bibfnamefont{P.~G.} \bibnamefont{Kusalik}},
  \bibnamefont{and} \bibinfo{author}{\bibfnamefont{G.}~\bibnamefont{Ciccotti}},
  \bibinfo{journal}{J. Chem. Phys.} \textbf{\bibinfo{volume}{142}},
  \bibinfo{pages}{244503} (\bibinfo{year}{2015}).

\bibitem[{\citenamefont{Singh and
  M{\"u}ller-Plathe}(2014)}]{singh2014characterization}
\bibinfo{author}{\bibfnamefont{J.~K.} \bibnamefont{Singh}} \bibnamefont{and}
  \bibinfo{author}{\bibfnamefont{F.}~\bibnamefont{M{\"u}ller-Plathe}},
  \bibinfo{journal}{Appl. Phys. Lett.} \textbf{\bibinfo{volume}{104}},
  \bibinfo{pages}{021603} (\bibinfo{year}{2014}).

\bibitem[{\citenamefont{Ni and Dijkstra}(2013)}]{ni2013effect}
\bibinfo{author}{\bibfnamefont{R.}~\bibnamefont{Ni}} \bibnamefont{and}
  \bibinfo{author}{\bibfnamefont{M.}~\bibnamefont{Dijkstra}},
  \bibinfo{journal}{Soft Matter} \textbf{\bibinfo{volume}{9}},
  \bibinfo{pages}{365} (\bibinfo{year}{2013}).

\bibitem[{\citenamefont{Reinhardt and Doye}(2014)}]{reinhardt2014effects}
\bibinfo{author}{\bibfnamefont{A.}~\bibnamefont{Reinhardt}} \bibnamefont{and}
  \bibinfo{author}{\bibfnamefont{J.~P.} \bibnamefont{Doye}},
  \bibinfo{journal}{J. Chem. Phys.} \textbf{\bibinfo{volume}{141}},
  \bibinfo{pages}{084501} (\bibinfo{year}{2014}).

\bibitem[{\citenamefont{Ten~Wolde and Frenkel}(1998)}]{ten1998computer}
\bibinfo{author}{\bibfnamefont{P.~R.} \bibnamefont{Ten~Wolde}}
  \bibnamefont{and} \bibinfo{author}{\bibfnamefont{D.}~\bibnamefont{Frenkel}},
  \bibinfo{journal}{J. Chem. Phys.} \textbf{\bibinfo{volume}{109}},
  \bibinfo{pages}{9901} (\bibinfo{year}{1998}).

\bibitem[{\citenamefont{Chen et~al.}(2001)\citenamefont{Chen, Siepmann, Oh, and
  Klein}}]{chen2001aggregation}
\bibinfo{author}{\bibfnamefont{B.}~\bibnamefont{Chen}},
  \bibinfo{author}{\bibfnamefont{J.~I.} \bibnamefont{Siepmann}},
  \bibinfo{author}{\bibfnamefont{K.~J.} \bibnamefont{Oh}}, \bibnamefont{and}
  \bibinfo{author}{\bibfnamefont{M.~L.} \bibnamefont{Klein}},
  \bibinfo{journal}{J. Chem. Phys.} \textbf{\bibinfo{volume}{115}},
  \bibinfo{pages}{10903} (\bibinfo{year}{2001}).

\bibitem[{\citenamefont{Oh and Zeng}(1999)}]{oh1999formation}
\bibinfo{author}{\bibfnamefont{K.}~\bibnamefont{Oh}} \bibnamefont{and}
  \bibinfo{author}{\bibfnamefont{X.~C.} \bibnamefont{Zeng}},
  \bibinfo{journal}{J. Chem. Phys.} \textbf{\bibinfo{volume}{110}},
  \bibinfo{pages}{4471} (\bibinfo{year}{1999}).

\bibitem[{\citenamefont{Chen et~al.}(2002)\citenamefont{Chen, Siepmann, Oh, and
  Klein}}]{chen2002simulating}
\bibinfo{author}{\bibfnamefont{B.}~\bibnamefont{Chen}},
  \bibinfo{author}{\bibfnamefont{J.~I.} \bibnamefont{Siepmann}},
  \bibinfo{author}{\bibfnamefont{K.~J.} \bibnamefont{Oh}}, \bibnamefont{and}
  \bibinfo{author}{\bibfnamefont{M.~L.} \bibnamefont{Klein}},
  \bibinfo{journal}{J. Chem. Phys.} \textbf{\bibinfo{volume}{116}},
  \bibinfo{pages}{4317} (\bibinfo{year}{2002}).

\bibitem[{\citenamefont{Zhukhovitskii}(1995)}]{zhukhovitskii1995molecular}
\bibinfo{author}{\bibfnamefont{D.}~\bibnamefont{Zhukhovitskii}},
  \bibinfo{journal}{J. Chem. Phys.} \textbf{\bibinfo{volume}{103}},
  \bibinfo{pages}{9401} (\bibinfo{year}{1995}).

\bibitem[{\citenamefont{Nishi et~al.}(2015)\citenamefont{Nishi, Inoue, and
  Matsumura}}]{nishi2015molecular}
\bibinfo{author}{\bibfnamefont{K.}~\bibnamefont{Nishi}},
  \bibinfo{author}{\bibfnamefont{S.}~\bibnamefont{Inoue}}, \bibnamefont{and}
  \bibinfo{author}{\bibfnamefont{Y.}~\bibnamefont{Matsumura}},
  \bibinfo{journal}{Chem. Phys. Lett.} \textbf{\bibinfo{volume}{634}},
  \bibinfo{pages}{194} (\bibinfo{year}{2015}).

\bibitem[{\citenamefont{Lupi et~al.}(2016)\citenamefont{Lupi, Peters, and
  Molinero}}]{lupi2016pre}
\bibinfo{author}{\bibfnamefont{L.}~\bibnamefont{Lupi}},
  \bibinfo{author}{\bibfnamefont{B.}~\bibnamefont{Peters}}, \bibnamefont{and}
  \bibinfo{author}{\bibfnamefont{V.}~\bibnamefont{Molinero}},
  \bibinfo{journal}{J. Chem. Phys.} \textbf{\bibinfo{volume}{145}},
  \bibinfo{pages}{211910} (\bibinfo{year}{2016}).

\bibitem[{\citenamefont{Santiso and Trout}(2015)}]{santiso2015general}
\bibinfo{author}{\bibfnamefont{E.~E.} \bibnamefont{Santiso}} \bibnamefont{and}
  \bibinfo{author}{\bibfnamefont{B.~L.} \bibnamefont{Trout}},
  \bibinfo{journal}{J. Chem. Phys.} \textbf{\bibinfo{volume}{143}},
  \bibinfo{pages}{174109} (\bibinfo{year}{2015}).

\bibitem[{\citenamefont{Berryman et~al.}(2016)\citenamefont{Berryman, Anwar,
  Dorosz, and Schilling}}]{berryman2016early}
\bibinfo{author}{\bibfnamefont{J.~T.} \bibnamefont{Berryman}},
  \bibinfo{author}{\bibfnamefont{M.}~\bibnamefont{Anwar}},
  \bibinfo{author}{\bibfnamefont{S.}~\bibnamefont{Dorosz}}, \bibnamefont{and}
  \bibinfo{author}{\bibfnamefont{T.}~\bibnamefont{Schilling}},
  \bibinfo{journal}{J. Chem. Phys.} \textbf{\bibinfo{volume}{145}},
  \bibinfo{pages}{211901} (\bibinfo{year}{2016}).

\bibitem[{\citenamefont{Zimmermann et~al.}(2015)\citenamefont{Zimmermann,
  Vorselaars, Quigley, and Peters}}]{zimmermann2015nucleation}
\bibinfo{author}{\bibfnamefont{N.~E.} \bibnamefont{Zimmermann}},
  \bibinfo{author}{\bibfnamefont{B.}~\bibnamefont{Vorselaars}},
  \bibinfo{author}{\bibfnamefont{D.}~\bibnamefont{Quigley}}, \bibnamefont{and}
  \bibinfo{author}{\bibfnamefont{B.}~\bibnamefont{Peters}},
  \bibinfo{journal}{J. Am. Chem. Soc.} \textbf{\bibinfo{volume}{137}},
  \bibinfo{pages}{13352} (\bibinfo{year}{2015}).

\bibitem[{\citenamefont{Lam et~al.}(2015)\citenamefont{Lam, Amans, Dujardin,
  Ledoux, and Allouche}}]{lam2015atomistic}
\bibinfo{author}{\bibfnamefont{J.}~\bibnamefont{Lam}},
  \bibinfo{author}{\bibfnamefont{D.}~\bibnamefont{Amans}},
  \bibinfo{author}{\bibfnamefont{C.}~\bibnamefont{Dujardin}},
  \bibinfo{author}{\bibfnamefont{G.}~\bibnamefont{Ledoux}}, \bibnamefont{and}
  \bibinfo{author}{\bibfnamefont{A.-R.} \bibnamefont{Allouche}},
  \bibinfo{journal}{J. Phys. Chem. A} \textbf{\bibinfo{volume}{119}},
  \bibinfo{pages}{8944} (\bibinfo{year}{2015}).

\bibitem[{\citenamefont{Kratzer and Arnold}(2015)}]{kratzer2015two}
\bibinfo{author}{\bibfnamefont{K.}~\bibnamefont{Kratzer}} \bibnamefont{and}
  \bibinfo{author}{\bibfnamefont{A.}~\bibnamefont{Arnold}},
  \bibinfo{journal}{Soft matter} \textbf{\bibinfo{volume}{11}},
  \bibinfo{pages}{2174} (\bibinfo{year}{2015}).

\bibitem[{\citenamefont{Bolhuis and Dellago}(2015)}]{bolhuis2015practical}
\bibinfo{author}{\bibfnamefont{P.}~\bibnamefont{Bolhuis}} \bibnamefont{and}
  \bibinfo{author}{\bibfnamefont{C.}~\bibnamefont{Dellago}},
  \bibinfo{journal}{Eur. Phys. J. Special Topics}
  \textbf{\bibinfo{volume}{224}}, \bibinfo{pages}{2409} (\bibinfo{year}{2015}).

\bibitem[{\citenamefont{Lau et~al.}(2015)\citenamefont{Lau, Hunt, M{\"u}ller,
  Jackson, and Ford}}]{lau2015water}
\bibinfo{author}{\bibfnamefont{G.~V.} \bibnamefont{Lau}},
  \bibinfo{author}{\bibfnamefont{P.~A.} \bibnamefont{Hunt}},
  \bibinfo{author}{\bibfnamefont{E.~A.} \bibnamefont{M{\"u}ller}},
  \bibinfo{author}{\bibfnamefont{G.}~\bibnamefont{Jackson}}, \bibnamefont{and}
  \bibinfo{author}{\bibfnamefont{I.~J.} \bibnamefont{Ford}},
  \bibinfo{journal}{J. Chem. Phys.} \textbf{\bibinfo{volume}{143}},
  \bibinfo{pages}{244709} (\bibinfo{year}{2015}).

\bibitem[{\citenamefont{Toxvaerd}(2016)}]{toxvaerd2016nucleation}
\bibinfo{author}{\bibfnamefont{S.}~\bibnamefont{Toxvaerd}},
  \bibinfo{journal}{J. Chem. Phys.} \textbf{\bibinfo{volume}{144}},
  \bibinfo{pages}{164502} (\bibinfo{year}{2016}).

\bibitem[{\citenamefont{Tanaka et~al.}(2005)\citenamefont{Tanaka, Kawamura,
  Tanaka, and Nakazawa}}]{tanaka2005tests}
\bibinfo{author}{\bibfnamefont{K.~K.} \bibnamefont{Tanaka}},
  \bibinfo{author}{\bibfnamefont{K.}~\bibnamefont{Kawamura}},
  \bibinfo{author}{\bibfnamefont{H.}~\bibnamefont{Tanaka}}, \bibnamefont{and}
  \bibinfo{author}{\bibfnamefont{K.}~\bibnamefont{Nakazawa}},
  \bibinfo{journal}{J. Chem. Phys.} \textbf{\bibinfo{volume}{122}},
  \bibinfo{pages}{184514} (\bibinfo{year}{2005}).

\bibitem[{\citenamefont{Kraska}(2006)}]{kraska2006molecular}
\bibinfo{author}{\bibfnamefont{T.}~\bibnamefont{Kraska}}, \bibinfo{journal}{J.
  Chem. Phys.} \textbf{\bibinfo{volume}{124}}, \bibinfo{pages}{054507}
  (\bibinfo{year}{2006}).

\bibitem[{\citenamefont{Oh and Zeng}(2000)}]{oh2000small}
\bibinfo{author}{\bibfnamefont{K.}~\bibnamefont{Oh}} \bibnamefont{and}
  \bibinfo{author}{\bibfnamefont{X.~C.} \bibnamefont{Zeng}},
  \bibinfo{journal}{J. Chem. Phys.} \textbf{\bibinfo{volume}{112}},
  \bibinfo{pages}{294} (\bibinfo{year}{2000}).

\bibitem[{\citenamefont{Senger et~al.}(1999)\citenamefont{Senger, Schaaf,
  Corti, Bowles, Pointu, Voegel, and Reiss}}]{senger1999molecular}
\bibinfo{author}{\bibfnamefont{B.}~\bibnamefont{Senger}},
  \bibinfo{author}{\bibfnamefont{P.}~\bibnamefont{Schaaf}},
  \bibinfo{author}{\bibfnamefont{D.}~\bibnamefont{Corti}},
  \bibinfo{author}{\bibfnamefont{R.}~\bibnamefont{Bowles}},
  \bibinfo{author}{\bibfnamefont{D.}~\bibnamefont{Pointu}},
  \bibinfo{author}{\bibfnamefont{J.-C.} \bibnamefont{Voegel}},
  \bibnamefont{and} \bibinfo{author}{\bibfnamefont{H.}~\bibnamefont{Reiss}},
  \bibinfo{journal}{J. Chem. Phys.} \textbf{\bibinfo{volume}{110}},
  \bibinfo{pages}{6438} (\bibinfo{year}{1999}).

\bibitem[{\citenamefont{Kulmala and Laaksonen}(1990)}]{kulmala1990binary}
\bibinfo{author}{\bibfnamefont{M.}~\bibnamefont{Kulmala}} \bibnamefont{and}
  \bibinfo{author}{\bibfnamefont{A.}~\bibnamefont{Laaksonen}},
  \bibinfo{journal}{J. Chem. Phys..} \textbf{\bibinfo{volume}{93}},
  \bibinfo{pages}{696} (\bibinfo{year}{1990}).

\bibitem[{\citenamefont{Zeng and Oxtoby}(1991{\natexlab{b}})}]{zeng1991binary}
\bibinfo{author}{\bibfnamefont{X.~C.} \bibnamefont{Zeng}} \bibnamefont{and}
  \bibinfo{author}{\bibfnamefont{D.}~\bibnamefont{Oxtoby}},
  \bibinfo{journal}{J. Chem. Phys.} \textbf{\bibinfo{volume}{95}},
  \bibinfo{pages}{5940} (\bibinfo{year}{1991}{\natexlab{b}}).

\bibitem[{\citenamefont{Oxtoby and Kashchiev}(1994)}]{oxtoby1994general}
\bibinfo{author}{\bibfnamefont{D.~W.} \bibnamefont{Oxtoby}} \bibnamefont{and}
  \bibinfo{author}{\bibfnamefont{D.}~\bibnamefont{Kashchiev}},
  \bibinfo{journal}{J. Chem. Phys.} \textbf{\bibinfo{volume}{100}},
  \bibinfo{pages}{7665} (\bibinfo{year}{1994}).

\bibitem[{\citenamefont{Napari and Laaksonen}(1999)}]{napari1999gas}
\bibinfo{author}{\bibfnamefont{I.}~\bibnamefont{Napari}} \bibnamefont{and}
  \bibinfo{author}{\bibfnamefont{A.}~\bibnamefont{Laaksonen}},
  \bibinfo{journal}{J. Chem. Phys.} \textbf{\bibinfo{volume}{111}},
  \bibinfo{pages}{5485} (\bibinfo{year}{1999}).

\bibitem[{\citenamefont{Jaecker-Voirol and
  Mirabel}(1988)}]{jaecker1988nucleation}
\bibinfo{author}{\bibfnamefont{A.}~\bibnamefont{Jaecker-Voirol}}
  \bibnamefont{and} \bibinfo{author}{\bibfnamefont{P.}~\bibnamefont{Mirabel}},
  \bibinfo{journal}{J. Phys. Chem.} \textbf{\bibinfo{volume}{92}},
  \bibinfo{pages}{3518} (\bibinfo{year}{1988}).

\bibitem[{\citenamefont{Talanquer and
  Oxtoby}(1995{\natexlab{b}})}]{talanquer1995nucleation}
\bibinfo{author}{\bibfnamefont{V.}~\bibnamefont{Talanquer}} \bibnamefont{and}
  \bibinfo{author}{\bibfnamefont{D.~W.} \bibnamefont{Oxtoby}},
  \bibinfo{journal}{J. Chem. Phys.} \textbf{\bibinfo{volume}{102}},
  \bibinfo{pages}{2156} (\bibinfo{year}{1995}{\natexlab{b}}).

\bibitem[{\citenamefont{ten Wolde and Frenkel}(1998)}]{ten1998numerical}
\bibinfo{author}{\bibfnamefont{P.~R.} \bibnamefont{ten Wolde}}
  \bibnamefont{and} \bibinfo{author}{\bibfnamefont{D.}~\bibnamefont{Frenkel}},
  \bibinfo{journal}{J. Chem. Phys.} \textbf{\bibinfo{volume}{109}},
  \bibinfo{pages}{9919} (\bibinfo{year}{1998}).

\bibitem[{\citenamefont{Laaksonen and Oxtoby}(1995)}]{laaksonen1995gas}
\bibinfo{author}{\bibfnamefont{A.}~\bibnamefont{Laaksonen}} \bibnamefont{and}
  \bibinfo{author}{\bibfnamefont{D.~W.} \bibnamefont{Oxtoby}},
  \bibinfo{journal}{J. Chem. Phys.} \textbf{\bibinfo{volume}{102}},
  \bibinfo{pages}{5803} (\bibinfo{year}{1995}).

\bibitem[{\citenamefont{Yoo et~al.}(2001)\citenamefont{Yoo, Oh, and
  Zeng}}]{yoo2001monte}
\bibinfo{author}{\bibfnamefont{S.}~\bibnamefont{Yoo}},
  \bibinfo{author}{\bibfnamefont{K.}~\bibnamefont{Oh}}, \bibnamefont{and}
  \bibinfo{author}{\bibfnamefont{X.~C.} \bibnamefont{Zeng}},
  \bibinfo{journal}{J. Chem. Phys.} \textbf{\bibinfo{volume}{115}},
  \bibinfo{pages}{8518} (\bibinfo{year}{2001}).

\bibitem[{\citenamefont{Napari and Laaksonen}(2000)}]{napari2000surfactant}
\bibinfo{author}{\bibfnamefont{I.}~\bibnamefont{Napari}} \bibnamefont{and}
  \bibinfo{author}{\bibfnamefont{A.}~\bibnamefont{Laaksonen}},
  \bibinfo{journal}{Phys. Rev. Lett.} \textbf{\bibinfo{volume}{84}},
  \bibinfo{pages}{2184} (\bibinfo{year}{2000}).

\bibitem[{\citenamefont{Braun et~al.}(2014)\citenamefont{Braun, Kalikmanov, and
  Kraska}}]{braun2014molecular}
\bibinfo{author}{\bibfnamefont{S.}~\bibnamefont{Braun}},
  \bibinfo{author}{\bibfnamefont{V.}~\bibnamefont{Kalikmanov}},
  \bibnamefont{and} \bibinfo{author}{\bibfnamefont{T.}~\bibnamefont{Kraska}},
  \bibinfo{journal}{J. Chem. Phys.} \textbf{\bibinfo{volume}{140}},
  \bibinfo{pages}{124305} (\bibinfo{year}{2014}).

\bibitem[{\citenamefont{Shimizu and Tanaka}(2015)}]{shimizu2015novel}
\bibinfo{author}{\bibfnamefont{R.}~\bibnamefont{Shimizu}} \bibnamefont{and}
  \bibinfo{author}{\bibfnamefont{H.}~\bibnamefont{Tanaka}},
  \bibinfo{journal}{Nature Commun.} \textbf{\bibinfo{volume}{6}},
  \bibinfo{pages}{7407} (\bibinfo{year}{2015}).

\bibitem[{\citenamefont{Pinho et~al.}(2014)\citenamefont{Pinho, Girardon,
  Bazer-Bachi, Bergeot, Marre, and Aymonier}}]{pinho2014microfluidic}
\bibinfo{author}{\bibfnamefont{B.}~\bibnamefont{Pinho}},
  \bibinfo{author}{\bibfnamefont{S.}~\bibnamefont{Girardon}},
  \bibinfo{author}{\bibfnamefont{F.}~\bibnamefont{Bazer-Bachi}},
  \bibinfo{author}{\bibfnamefont{G.}~\bibnamefont{Bergeot}},
  \bibinfo{author}{\bibfnamefont{S.}~\bibnamefont{Marre}}, \bibnamefont{and}
  \bibinfo{author}{\bibfnamefont{C.}~\bibnamefont{Aymonier}},
  \bibinfo{journal}{Lab on a Chip} \textbf{\bibinfo{volume}{14}},
  \bibinfo{pages}{3843} (\bibinfo{year}{2014}).

\bibitem[{\citenamefont{Gao et~al.}(2014)\citenamefont{Gao, Fu, Xie, Su, and
  Wang}}]{gao2014}
\bibinfo{author}{\bibfnamefont{X.}~\bibnamefont{Gao}},
  \bibinfo{author}{\bibfnamefont{D.}~\bibnamefont{Fu}},
  \bibinfo{author}{\bibfnamefont{B.}~\bibnamefont{Xie}},
  \bibinfo{author}{\bibfnamefont{Y.}~\bibnamefont{Su}}, \bibnamefont{and}
  \bibinfo{author}{\bibfnamefont{D.}~\bibnamefont{Wang}}, \bibinfo{journal}{J.
  Phys. Chem. B} \textbf{\bibinfo{volume}{118}}, \bibinfo{pages}{12549}
  (\bibinfo{year}{2014}).

\bibitem[{\citenamefont{Alekseechkin}(2015)}]{alekseechkin2015thermodynamics}
\bibinfo{author}{\bibfnamefont{N.~V.} \bibnamefont{Alekseechkin}},
  \bibinfo{journal}{J. Chem. Phys.} \textbf{\bibinfo{volume}{143}},
  \bibinfo{pages}{054502} (\bibinfo{year}{2015}).

\bibitem[{\citenamefont{Watson et~al.}(2011)\citenamefont{Watson, Nguelo,
  Desgranges, and Delhommelle}}]{watson2011crystal}
\bibinfo{author}{\bibfnamefont{K.~D.} \bibnamefont{Watson}},
  \bibinfo{author}{\bibfnamefont{S.~T.} \bibnamefont{Nguelo}},
  \bibinfo{author}{\bibfnamefont{C.}~\bibnamefont{Desgranges}},
  \bibnamefont{and}
  \bibinfo{author}{\bibfnamefont{J.}~\bibnamefont{Delhommelle}},
  \bibinfo{journal}{CrystEngComm} \textbf{\bibinfo{volume}{13}},
  \bibinfo{pages}{1132} (\bibinfo{year}{2011}).

\bibitem[{\citenamefont{Desgranges and
  Delhommelle}(2014{\natexlab{a}})}]{desgranges2014unraveling}
\bibinfo{author}{\bibfnamefont{C.}~\bibnamefont{Desgranges}} \bibnamefont{and}
  \bibinfo{author}{\bibfnamefont{J.}~\bibnamefont{Delhommelle}},
  \bibinfo{journal}{J. Am. Chem. Soc.} \textbf{\bibinfo{volume}{136}},
  \bibinfo{pages}{8145} (\bibinfo{year}{2014}{\natexlab{a}}).

\bibitem[{\citenamefont{Desgranges and
  Delhommelle}(2016{\natexlab{a}})}]{muVTS1}
\bibinfo{author}{\bibfnamefont{C.}~\bibnamefont{Desgranges}} \bibnamefont{and}
  \bibinfo{author}{\bibfnamefont{J.}~\bibnamefont{Delhommelle}},
  \bibinfo{journal}{J. Chem. Phys. - Part I (accepted for publication)}
  (\bibinfo{year}{2016}{\natexlab{a}}).

\bibitem[{\citenamefont{McGraw and Laaksonen}(1996)}]{mcgraw1996scaling}
\bibinfo{author}{\bibfnamefont{R.}~\bibnamefont{McGraw}} \bibnamefont{and}
  \bibinfo{author}{\bibfnamefont{A.}~\bibnamefont{Laaksonen}},
  \bibinfo{journal}{Phys. Rev. Lett.} \textbf{\bibinfo{volume}{76}},
  \bibinfo{pages}{2754} (\bibinfo{year}{1996}).

\bibitem[{\citenamefont{Torrie and Valleau}(1977)}]{torrie1977nonphysical}
\bibinfo{author}{\bibfnamefont{G.~M.} \bibnamefont{Torrie}} \bibnamefont{and}
  \bibinfo{author}{\bibfnamefont{J.~P.} \bibnamefont{Valleau}},
  \bibinfo{journal}{J. Comput, Phys.} \textbf{\bibinfo{volume}{23}},
  \bibinfo{pages}{187} (\bibinfo{year}{1977}).

\bibitem[{\citenamefont{Allen and Tildesley}(1987)}]{Allen}
\bibinfo{author}{\bibfnamefont{M.~P.} \bibnamefont{Allen}} \bibnamefont{and}
  \bibinfo{author}{\bibfnamefont{D.~J.} \bibnamefont{Tildesley}},
  \emph{\bibinfo{title}{Computer Simulation of Liquids}}
  (\bibinfo{publisher}{Clarendon Press, Oxford}, \bibinfo{year}{1987}).

\bibitem[{\citenamefont{Desgranges and Delhommelle}(2009)}]{Au}
\bibinfo{author}{\bibfnamefont{C.}~\bibnamefont{Desgranges}} \bibnamefont{and}
  \bibinfo{author}{\bibfnamefont{J.}~\bibnamefont{Delhommelle}},
  \bibinfo{journal}{J. Phys. Chem. C} \textbf{\bibinfo{volume}{113}},
  \bibinfo{pages}{3607} (\bibinfo{year}{2009}).

\bibitem[{\citenamefont{Desgranges and Delhommelle}(2007)}]{Alu2}
\bibinfo{author}{\bibfnamefont{C.}~\bibnamefont{Desgranges}} \bibnamefont{and}
  \bibinfo{author}{\bibfnamefont{J.}~\bibnamefont{Delhommelle}},
  \bibinfo{journal}{J. Chem. Phys.} \textbf{\bibinfo{volume}{127}},
  \bibinfo{pages}{144509} (\bibinfo{year}{2007}).

\bibitem[{\citenamefont{Vrabec et~al.}(2001)\citenamefont{Vrabec, Stoll, and
  Hasse}}]{vrabec2001set}
\bibinfo{author}{\bibfnamefont{J.}~\bibnamefont{Vrabec}},
  \bibinfo{author}{\bibfnamefont{J.}~\bibnamefont{Stoll}}, \bibnamefont{and}
  \bibinfo{author}{\bibfnamefont{H.}~\bibnamefont{Hasse}}, \bibinfo{journal}{J.
  Phys. Chem. B} \textbf{\bibinfo{volume}{105}}, \bibinfo{pages}{12126}
  (\bibinfo{year}{2001}).

\bibitem[{\citenamefont{Potoff et~al.}(1999)\citenamefont{Potoff, Errington,
  and Panagiotopoulos}}]{Potoff}
\bibinfo{author}{\bibfnamefont{J.~J.} \bibnamefont{Potoff}},
  \bibinfo{author}{\bibfnamefont{J.~R.} \bibnamefont{Errington}},
  \bibnamefont{and} \bibinfo{author}{\bibfnamefont{A.~Z.}
  \bibnamefont{Panagiotopoulos}}, \bibinfo{journal}{Mol. Phys.}
  \textbf{\bibinfo{volume}{97}}, \bibinfo{pages}{1073} (\bibinfo{year}{1999}).

\bibitem[{\citenamefont{Errington and
  Panagiotopoulos}(1999{\natexlab{a}})}]{Errington1}
\bibinfo{author}{\bibfnamefont{J.~R.} \bibnamefont{Errington}}
  \bibnamefont{and} \bibinfo{author}{\bibfnamefont{A.~Z.}
  \bibnamefont{Panagiotopoulos}}, \bibinfo{journal}{J. Phys. Chem. B}
  \textbf{\bibinfo{volume}{103}}, \bibinfo{pages}{6314}
  (\bibinfo{year}{1999}{\natexlab{a}}).

\bibitem[{\citenamefont{Errington and
  Panagiotopoulos}(1999{\natexlab{b}})}]{Errington2}
\bibinfo{author}{\bibfnamefont{J.~R.} \bibnamefont{Errington}}
  \bibnamefont{and} \bibinfo{author}{\bibfnamefont{A.~Z.}
  \bibnamefont{Panagiotopoulos}}, \bibinfo{journal}{J. Chem. Phys.}
  \textbf{\bibinfo{volume}{111}}, \bibinfo{pages}{9731}
  (\bibinfo{year}{1999}{\natexlab{b}}).

\bibitem[{\citenamefont{Errington}(2003)}]{Errington3}
\bibinfo{author}{\bibfnamefont{J.~R.} \bibnamefont{Errington}},
  \bibinfo{journal}{J. Chem. Phys.} \textbf{\bibinfo{volume}{118}},
  \bibinfo{pages}{9915} (\bibinfo{year}{2003}).

\bibitem[{\citenamefont{Desgranges and
  Delhommelle}(2014{\natexlab{b}})}]{PartIII}
\bibinfo{author}{\bibfnamefont{C.}~\bibnamefont{Desgranges}} \bibnamefont{and}
  \bibinfo{author}{\bibfnamefont{J.}~\bibnamefont{Delhommelle}},
  \bibinfo{journal}{J. Chem. Phys.} \textbf{\bibinfo{volume}{140}},
  \bibinfo{pages}{104109} (\bibinfo{year}{2014}{\natexlab{b}}).

\bibitem[{\citenamefont{Desgranges and
  Delhommelle}(2012{\natexlab{a}})}]{PartI}
\bibinfo{author}{\bibfnamefont{C.}~\bibnamefont{Desgranges}} \bibnamefont{and}
  \bibinfo{author}{\bibfnamefont{J.}~\bibnamefont{Delhommelle}},
  \bibinfo{journal}{J. Chem. Phys.} \textbf{\bibinfo{volume}{136}},
  \bibinfo{pages}{184107} (\bibinfo{year}{2012}{\natexlab{a}}).

\bibitem[{\citenamefont{Desgranges and
  Delhommelle}(2012{\natexlab{b}})}]{PartII}
\bibinfo{author}{\bibfnamefont{C.}~\bibnamefont{Desgranges}} \bibnamefont{and}
  \bibinfo{author}{\bibfnamefont{J.}~\bibnamefont{Delhommelle}},
  \bibinfo{journal}{J. Chem. Phys.} \textbf{\bibinfo{volume}{136}},
  \bibinfo{pages}{184108} (\bibinfo{year}{2012}{\natexlab{b}}).

\bibitem[{\citenamefont{Desgranges and
  Delhommelle}(2016{\natexlab{b}})}]{PartIV}
\bibinfo{author}{\bibfnamefont{C.}~\bibnamefont{Desgranges}} \bibnamefont{and}
  \bibinfo{author}{\bibfnamefont{J.}~\bibnamefont{Delhommelle}},
  \bibinfo{journal}{J. Chem. Phys.} \textbf{\bibinfo{volume}{144}},
  \bibinfo{pages}{124510} (\bibinfo{year}{2016}{\natexlab{b}}).

\bibitem[{\citenamefont{Gazenm$\ddot{\mathrm{u}}$ller and Camp}(2007)}]{Camp}
\bibinfo{author}{\bibfnamefont{G.}~\bibnamefont{Gazenm$\ddot{\mathrm{u}}$ller}}
  \bibnamefont{and} \bibinfo{author}{\bibfnamefont{P.~J.} \bibnamefont{Camp}},
  \bibinfo{journal}{J. Chem. Phys.} \textbf{\bibinfo{volume}{127}},
  \bibinfo{pages}{154504} (\bibinfo{year}{2007}).

\bibitem[{\citenamefont{Liu et~al.}(2011)\citenamefont{Liu, Panagiotopoulos,
  and Debenedetti}}]{Pana}
\bibinfo{author}{\bibfnamefont{Y.}~\bibnamefont{Liu}},
  \bibinfo{author}{\bibfnamefont{A.~Z.} \bibnamefont{Panagiotopoulos}},
  \bibnamefont{and} \bibinfo{author}{\bibfnamefont{P.~G.}
  \bibnamefont{Debenedetti}}, \bibinfo{journal}{J. Phys. Chem. B}
  \textbf{\bibinfo{volume}{115}}, \bibinfo{pages}{6629} (\bibinfo{year}{2011}).

\bibitem[{\citenamefont{Nezbeda and Kolafa}(1991)}]{nezbeda1991new}
\bibinfo{author}{\bibfnamefont{I.}~\bibnamefont{Nezbeda}} \bibnamefont{and}
  \bibinfo{author}{\bibfnamefont{J.}~\bibnamefont{Kolafa}},
  \bibinfo{journal}{Molec. Simul.} \textbf{\bibinfo{volume}{5}},
  \bibinfo{pages}{391} (\bibinfo{year}{1991}).

\bibitem[{\citenamefont{Singh and Errington}(2006)}]{Singh}
\bibinfo{author}{\bibfnamefont{J.~K.} \bibnamefont{Singh}} \bibnamefont{and}
  \bibinfo{author}{\bibfnamefont{J.~R.} \bibnamefont{Errington}},
  \bibinfo{journal}{J. Phys. Chem. B} \textbf{\bibinfo{volume}{110}},
  \bibinfo{pages}{1369} (\bibinfo{year}{2006}).

\bibitem[{\citenamefont{Rai et~al.}(2007)\citenamefont{Rai, Siepmann, Schultz,
  and Ross}}]{rai2007pressure}
\bibinfo{author}{\bibfnamefont{N.}~\bibnamefont{Rai}},
  \bibinfo{author}{\bibfnamefont{J.~I.} \bibnamefont{Siepmann}},
  \bibinfo{author}{\bibfnamefont{N.~E.} \bibnamefont{Schultz}},
  \bibnamefont{and} \bibinfo{author}{\bibfnamefont{R.~B.} \bibnamefont{Ross}},
  \bibinfo{journal}{J. Phys. Chem. C} \textbf{\bibinfo{volume}{111}},
  \bibinfo{pages}{15634} (\bibinfo{year}{2007}).

\bibitem[{\citenamefont{Rane et~al.}(2013)\citenamefont{Rane, Murali, and
  Errington}}]{Rane}
\bibinfo{author}{\bibfnamefont{K.~S.} \bibnamefont{Rane}},
  \bibinfo{author}{\bibfnamefont{S.}~\bibnamefont{Murali}}, \bibnamefont{and}
  \bibinfo{author}{\bibfnamefont{J.~R.} \bibnamefont{Errington}},
  \bibinfo{journal}{J. Chem. Theory Comput.} \textbf{\bibinfo{volume}{9}},
  \bibinfo{pages}{2552} (\bibinfo{year}{2013}).

\bibitem[{\citenamefont{Escobedo and de~Pablo}(1996)}]{expanded}
\bibinfo{author}{\bibfnamefont{F.}~\bibnamefont{Escobedo}} \bibnamefont{and}
  \bibinfo{author}{\bibfnamefont{J.~J.} \bibnamefont{de~Pablo}},
  \bibinfo{journal}{J. Chem. Phys.} \textbf{\bibinfo{volume}{105}},
  \bibinfo{pages}{4391} (\bibinfo{year}{1996}).

\bibitem[{\citenamefont{Shi and Maginn}(2008)}]{Shi1}
\bibinfo{author}{\bibfnamefont{W.}~\bibnamefont{Shi}} \bibnamefont{and}
  \bibinfo{author}{\bibfnamefont{E.~J.} \bibnamefont{Maginn}},
  \bibinfo{journal}{J. Comp. Chem.} \textbf{\bibinfo{volume}{29}},
  \bibinfo{pages}{2520} (\bibinfo{year}{2008}).

\bibitem[{\citenamefont{Eslami and
  M{\"u}ller-Plathe}(2007)}]{eslami2007molecular}
\bibinfo{author}{\bibfnamefont{H.}~\bibnamefont{Eslami}} \bibnamefont{and}
  \bibinfo{author}{\bibfnamefont{F.}~\bibnamefont{M{\"u}ller-Plathe}},
  \bibinfo{journal}{J. Comput. Chem.} \textbf{\bibinfo{volume}{28}},
  \bibinfo{pages}{1763} (\bibinfo{year}{2007}).

\bibitem[{\citenamefont{Vogt et~al.}(2001)\citenamefont{Vogt, Liapine,
  Kirchner, Dyson, Huber, Marcelli, and Sadus}}]{Vogt}
\bibinfo{author}{\bibfnamefont{P.~S.} \bibnamefont{Vogt}},
  \bibinfo{author}{\bibfnamefont{R.}~\bibnamefont{Liapine}},
  \bibinfo{author}{\bibfnamefont{B.}~\bibnamefont{Kirchner}},
  \bibinfo{author}{\bibfnamefont{A.~J.} \bibnamefont{Dyson}},
  \bibinfo{author}{\bibfnamefont{H.}~\bibnamefont{Huber}},
  \bibinfo{author}{\bibfnamefont{G.}~\bibnamefont{Marcelli}}, \bibnamefont{and}
  \bibinfo{author}{\bibfnamefont{R.~J.} \bibnamefont{Sadus}},
  \bibinfo{journal}{Phys. Chem. Chem. Phys.} \textbf{\bibinfo{volume}{3}},
  \bibinfo{pages}{1297} (\bibinfo{year}{2001}).

\bibitem[{\citenamefont{Widom}(1963)}]{widom1963some}
\bibinfo{author}{\bibfnamefont{B.}~\bibnamefont{Widom}}, \bibinfo{journal}{J.
  Chem. Phys.} \textbf{\bibinfo{volume}{39}}, \bibinfo{pages}{2808}
  (\bibinfo{year}{1963}).

\bibitem[{\citenamefont{Siepmann and Frenkel}(1992)}]{CBMC}
\bibinfo{author}{\bibfnamefont{J.}~\bibnamefont{Siepmann}} \bibnamefont{and}
  \bibinfo{author}{\bibfnamefont{D.}~\bibnamefont{Frenkel}},
  \bibinfo{journal}{Mol. Phys.} \textbf{\bibinfo{volume}{75}},
  \bibinfo{pages}{59} (\bibinfo{year}{1992}).

\bibitem[{\citenamefont{Reiss}(1950)}]{reiss1950kinetics}
\bibinfo{author}{\bibfnamefont{H.}~\bibnamefont{Reiss}}, \bibinfo{journal}{J.
  Chem. Phys.} \textbf{\bibinfo{volume}{18}}, \bibinfo{pages}{840}
  (\bibinfo{year}{1950}).

\bibitem[{\citenamefont{Wyslouzil and
  Seinfeld}(1992)}]{wyslouzil1992nonisothermal}
\bibinfo{author}{\bibfnamefont{B.}~\bibnamefont{Wyslouzil}} \bibnamefont{and}
  \bibinfo{author}{\bibfnamefont{J.}~\bibnamefont{Seinfeld}},
  \bibinfo{journal}{J. Chem. Phys.} \textbf{\bibinfo{volume}{97}},
  \bibinfo{pages}{2661} (\bibinfo{year}{1992}).

\bibitem[{\citenamefont{Wedekind et~al.}(2007)\citenamefont{Wedekind, Reguera,
  and Strey}}]{wedekind2007influence}
\bibinfo{author}{\bibfnamefont{J.}~\bibnamefont{Wedekind}},
  \bibinfo{author}{\bibfnamefont{D.}~\bibnamefont{Reguera}}, \bibnamefont{and}
  \bibinfo{author}{\bibfnamefont{R.}~\bibnamefont{Strey}}, \bibinfo{journal}{J.
  Chem. Phys.} \textbf{\bibinfo{volume}{127}}, \bibinfo{pages}{064501}
  (\bibinfo{year}{2007}).

\bibitem[{\citenamefont{Wilemski}(1987)}]{wilemski1987revised}
\bibinfo{author}{\bibfnamefont{G.}~\bibnamefont{Wilemski}},
  \bibinfo{journal}{Journal of Physical Chemistry}
  \textbf{\bibinfo{volume}{91}}, \bibinfo{pages}{2492} (\bibinfo{year}{1987}).

\bibitem[{\citenamefont{Napari et~al.}(1999)\citenamefont{Napari, Laaksonen,
  Talanquer, and Oxtoby}}]{napari1999density}
\bibinfo{author}{\bibfnamefont{I.}~\bibnamefont{Napari}},
  \bibinfo{author}{\bibfnamefont{A.}~\bibnamefont{Laaksonen}},
  \bibinfo{author}{\bibfnamefont{V.}~\bibnamefont{Talanquer}},
  \bibnamefont{and} \bibinfo{author}{\bibfnamefont{D.~W.}
  \bibnamefont{Oxtoby}}, \bibinfo{journal}{The Journal of chemical physics}
  \textbf{\bibinfo{volume}{110}}, \bibinfo{pages}{5906} (\bibinfo{year}{1999}).

\bibitem[{\citenamefont{Bennett}(1977)}]{bennett1977algorithms}
\bibinfo{author}{\bibfnamefont{C.~H.} \bibnamefont{Bennett}},
  \emph{\bibinfo{title}{Algorithms for chemical computations}}
  (\bibinfo{publisher}{ACS symposium Series, vol. 46, American Chemical
  Society, Washington DC}, \bibinfo{year}{1977}).

\bibitem[{\citenamefont{Chandler}(1978)}]{chandler1978statistical}
\bibinfo{author}{\bibfnamefont{D.}~\bibnamefont{Chandler}},
  \bibinfo{journal}{J. Chem. Phys.} \textbf{\bibinfo{volume}{68}},
  \bibinfo{pages}{2959} (\bibinfo{year}{1978}).

\bibitem[{\citenamefont{Carter et~al.}(1989)\citenamefont{Carter, Ciccotti,
  Hynes, and Kapral}}]{carter1989constrained}
\bibinfo{author}{\bibfnamefont{E.}~\bibnamefont{Carter}},
  \bibinfo{author}{\bibfnamefont{G.}~\bibnamefont{Ciccotti}},
  \bibinfo{author}{\bibfnamefont{J.~T.} \bibnamefont{Hynes}}, \bibnamefont{and}
  \bibinfo{author}{\bibfnamefont{R.}~\bibnamefont{Kapral}},
  \bibinfo{journal}{Chem. Phys. Lett.} \textbf{\bibinfo{volume}{156}},
  \bibinfo{pages}{472} (\bibinfo{year}{1989}).

\end{thebibliography}

\end{document}